\documentclass[aps,prd,twocolumn,showpacs,eqsecnum,superscriptaddress,floatfix,nofootinbib]{revtex4-1}
\usepackage{graphicx,amsmath,amssymb,epsfig,color,ulem,color}

\usepackage{hyperref} 
\hypersetup{
    unicode=true,          
    bookmarksnumbered=true,
    colorlinks=true,       
    linkcolor=red,          
    citecolor=[rgb]{0,.7,0},        
    filecolor=magenta,      
    urlcolor=cyan,           
    menucolor=blue,
    baseurl=http://www.arxiv.org/abs/,
    hyperfootnotes=true
}

\newcommand{\eqn}[2] {\begin{equation}\label{#1}{#2}\end{equation}}
\newcommand{\eno}[1] {\eqref{#1}}

\begin{document}

\begin{picture}(0,20)(0,0)
 \put(185,20){PUPT-2455}
\end{picture}

\title{Shooting String Holography of Jet Quenching at RHIC and LHC}

\author{Andrej Ficnar}
\email{aficnar@phys.columbia.edu}
\affiliation{Department of Physics, Columbia University, New York, NY 10027, USA}

\author{Steven S. Gubser}
\email{ssgubser@princeton.edu}
\affiliation{Joseph Henry Laboratories, Princeton University, Princeton, NJ 08544, USA}

\author{Miklos Gyulassy}
\email{gyulassy@phys.columbia.edu}
\affiliation{Department of Physics, Columbia University, New York, NY 10027, USA}

\date{October 2014}

\begin{abstract}
We derive a new formula for jet energy loss using finite endpoint momentum shooting strings initial conditions in SYM plasmas to overcome the difficulties of previous falling string holographic scenarios. We apply the new formula to compute the nuclear modification factor $R_{AA}$ and the elliptic flow parameter $v_2$ of light hadrons at RHIC and LHC. We show furthermore that Gauss-Bonnet quadratic curvature corrections to the $AdS_5$ geometry improve the agreement with the recent data.
\end{abstract}


\maketitle


\section{Introduction}
\label{intro}

An account of jet-quenching in AdS/CFT based on classical string trajectories has been developed in \cite{ggpr08,Chesler:2008wd} and a number of subsequent works, with a parallel line of development starting in \cite{Hatta:2008tx}.  The approach of \cite{ggpr08} was to model an energetic gluon as a string in $AdS_5$-Schwarzschild with both ends passing through the horizon.  Light quarks could be modeled analogously by having a string with one endpoint ending on a D7-brane in the bulk of $AdS_5$.  The complementary approach of \cite{Chesler:2008wd} was to model a light quark-antiquark pair by an initially pointlike open string created close to the boundary with endpoints that are free to fly apart.  In either case, the string extends in a direction parallel to the boundary as it falls toward the black hole horizon.  In all three works \cite{ggpr08,Hatta:2008tx,Chesler:2008wd}, it was found that the maximum distance that an energetic probe can travel for a fixed energy $E$ in a thermal $\mathcal{N}=4$ SYM plasma at a temperature $T$ scales as $\Delta x_{\rm max}\propto E^{1/3}T^{-4/3}$. The constant of proportionality is important for phenomenological applications as it determines the overall strength of jet quenching. 

To compute the observables such as the nuclear modification factor $R_{AA}$ and the elliptic flow parameter $v_2$ of light hadrons, we need to know the details of the instantaneous energy loss of light quarks. To tackle this problem, a general formula for computing the instantaneous energy loss in non-stationary string configurations was developed \cite{f12}, using methods related to the earlier work \cite{cjky08} but with somewhat different results. The application of this formula of \cite{f12} to the case of fallings strings requires a precise definition of the energy loss (i.e., roughly speaking, what part of the string is to be considered as the ``jet'' and what part as the thermalized energy to which the jet energy is being lost) and is susceptible to the details of the initial conditions. Nevertheless, studies have shown some rather universal qualitative features of the light quark energy loss \cite{f12}, including a modified Bragg peak and a seemingly linear path dependence which made it look very similar to the radiative pQCD energy loss. 

\begin{figure}
\centerline{\includegraphics[width=3.35in]{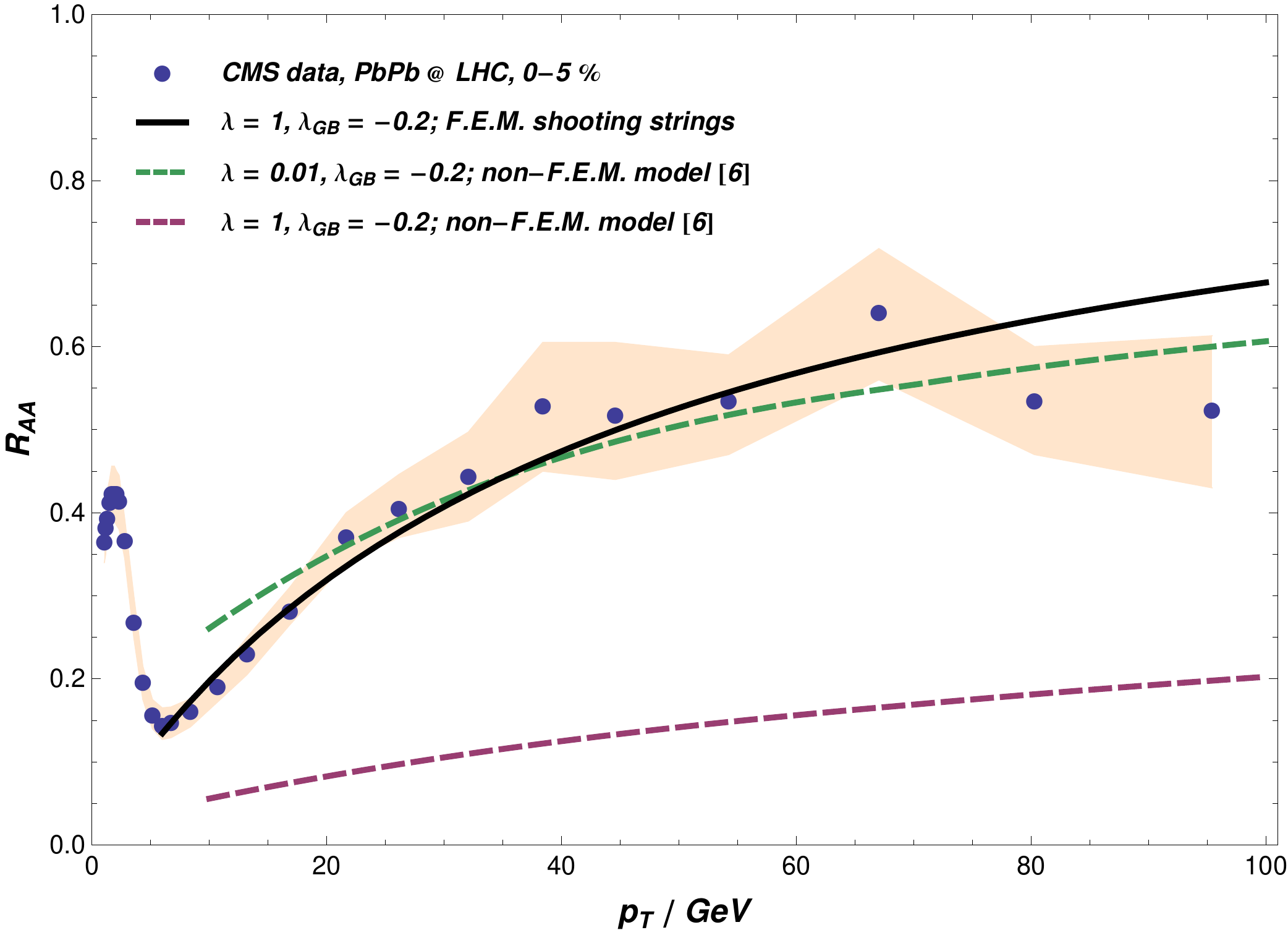}}
\caption{Model calculations of the nuclear suppression factor $R_{AA}$ of pions in central collisions at the LHC, compared to the CMS data \cite{cms12}. The dashed curves are based on the energy loss model inferred from a specific treatment of falling strings in \cite{fng12} that assumes no finite endpoint momentum (\textit{non-F.E.M.}). The solid line represents the $R_{AA}$ computed in the framework of the finite endpoint momentum strings, which we describe in this Letter. All three curves were computed with the higher derivative Gauss-Bonnet corrections to $AdS_5$.}
\label{IntroFig}
\end{figure}

Using these results for the stopping distance and the path dependence of the energy loss, one can perform the simplest constructions of the nuclear modification factor $R_{AA}$ \cite{fng12}\footnote{See Appendix \ref{fallGB} for calculations based on methods of \cite{fg13} that further support the claims and assumptions behind the $R_{AA}$ constructions in \cite{fng12}.}. The comparison with the LHC pion suppression data showed that, although it had the right qualitative structure, the overall magnitude was too low, indicating that the predicted jet quenching was too strong. Introducing higher derivative corrections to the $AdS_5$ showed substantial increase in $R_{AA}$, but this effect alone was not enough to get close to the data. This is all illustrated in Fig. \ref{IntroFig}, where the dashed red line shows how far below the data this best-case scenario ($\lambda=1$ and including the higher derivative corrections) is, and the dashed green curve shows how much we would effectively need to decrease the coupling (all the way to about $\lambda=0.01$) to come close to the data. 

A novel step towards addressing this as well as some other issues, and ultimately towards a more realistic description of energetic quarks, was taken recently by introduction of finite momentum at the endpoints of the string \cite{fg13}. Typically, standard boundary conditions for open strings require vanishing endpoint momentum, but if one thinks of the endpoints as representing energetic quarks and the string between them representing the color field they generate, then a string with most of its energy packed into its endpoints provides a more natural holographic dual to a pair of quarks that have undergone a hard scattering event. In this way, one also obtains a clear distinction between the energy in the hard probe and energy contained in the color fields surrounding it, hence offering a clear definition of the instantaneous jet energy loss that was missing in earlier accounts. Another important feature of this proposal is that the distance the finite momentum endpoints travel for a given energy is greater than in the previous treatments of the falling strings. In other words, quark jets that these strings represent are less quenched and hence offer a potentially better match with the experimental data. In this article, we aim to explore the phenomenological features of this proposal, some of which are summarized in Fig. \ref{IntroFig}, showing a good match with the data for $\lambda=1$ (solid black curve).

Our overall goal is to use the ideas of \cite{fg13} to provide a simultaneous fit to $R_{AA}$ from RHIC and LHC, with attention also to $v_2$ of hard probes.  We focus on a particular type of classical string motion, where the string endpoint starts near the horizon and then moves {\it upward} toward the boundary, carrying some amount of energy and momentum which is gradually bled off into the rest of the string during its rise.  These motions are termed finite-endpoint-momentum shooting strings, or ``F.E.M.\ shooting strings'' for short.

Modulo some assumptions, F.E.M.\ shooting strings lead to a concise and phenomenologically interesting formula for instantaneous energy loss, presented as Eq. \eno{eloss4}.  The energy loss depends explicitly on the 't Hooft coupling $\lambda$, and it receives corrections when higher derivative terms are included in the gravitational action.  We restrict attention to Gauss-Bonnet corrections, with parameter $\lambda_{GB}$.

In the following sections, we consider several different regimes of parameters, driven mostly by phenomenological considerations, but also by a desire to avoid small values of $\lambda$ which take us decisively outside the regime of validity of the supergravity approximation in AdS/CFT.  Other phenomenological parameters controlling the plasma equilibration time and the local evolution of temperature and radial velocity enter significantly into the discussion.  While it is challenging to simultaneously fit LHC and RHIC data, the choice $\lambda = 4$ and $\lambda_{GB} = -0.2$ puts our predictions in the ballpark of data {\it provided} we include a $10\%$ reduction of temperature at the LHC relative to straightforward expectations based on multiplicities.


\section{Energy loss}
\label{enloss}

In this section we will develop a phenomenologically usable form of the instantaneous energy loss $dE/dx$ based on the finite endpoint momentum framework. As shown in \cite{fg13}, a direct consequence of having finite endpoint momentum is that the trajectories of the endpoints are piecewise null geodesics in $AdS_5$ along which the endpoint momentum evolves according to equations that do not depend on the bulk shape of the string:
\eqn{eloss1}
{\frac{dE}{dx}=-\frac{\sqrt{\lambda}}{2\pi}\frac{\sqrt{f(z_*)}}{z^2}\,,}
where $\sqrt{\lambda}=L^2/\alpha'$ is the 't Hooft coupling, $f(z)=1-z^4/z_H^4$ (in this coordinate system the boundary is at $z=0$), $z_H=1/(\pi T)$ and $z_*$ is the minimal (inverse) radial coordinate the geodesic reaches and which hence completely determines the motion of the endpoint. See Appendix \ref{GeneralShoot} for a derivation of this formula in more general geometries. As mentioned before, considering endpoints as energetic quarks themselves and the string as the color field they generate, we will identify the rate at which the energy gets drained from the endpoint with the energy loss of an energetic quark. It is worth pointing out that \eno{eloss1} is a unique answer, independent of the initial conditions: it does not depend on the energy stored in the endpoint\footnote{In \cite{fg13}, $z_*$ was related to the initial energy of the endpoint $E_0$ by saying that the energy at the moment when the endpoint reaches the horizon must be zero. The reason behind this proposal was that this prescription maximized the stopping distance for a given energy, without the string performing a ``snap-back''. However, in general, it only matters that the string does not perform a snap-back within some phenomenologically relevant distance $L$ so for a given $z_*$ we can have a multitude of energies.} and it is a function of only the radial coordinate $z$ at which the endpoint is located, and weakly dependent on the $z_*$ of the geodesic along which the endpoint is moving.

To express $dE/dx$ as a function of $x$, we need to solve the null geodesic equation. Assuming that initially, at $x=0$, the endpoint is at $z=z_0$ going towards the boundary, we have:
\eqn{eloss2}
{x_{\rm geo}(z)=\frac{z_H^2}{z}\,_2F_1\left(\frac{1}{4},\frac{1}{2},\frac{5}{4},\frac{z_*^4}{z^4}\right)- \frac{z_H^2}{z_0}\,_2F_1\left(\frac{1}{4},\frac{1}{2},\frac{5}{4},\frac{z_*^4}{z_0^4}\right)\,,}
where ${_2F_1}$ is the ordinary hypergeometric function. One could now numerically invert this relation (for given $z_*$ and $z_0$) to obtain $z(x)$ and plug it in \eno{eloss1} to obtain $dE/dx$ as a function of $x$ which would result in a characteristic bell-shaped curve for energy loss \cite{fg13}. However, \eno{eloss2} has a particularly simple and universal form for small $z_*$:
\eqn{eloss3}
{x_{\rm geo}(z)=z_H^2 \left[ \left(\frac{1}{z}-\frac{1}{z_0}\right)+\mathcal{O}\left(\frac{z_*^4}{10 z^5},\frac{z_*^4}{10 z_0^5}\right) \right] \,.}
The reason we are interested in this expansion is phenomenological: from \eno{eloss1} we see that if we start at $z$ close to the boundary, the energy loss will be large, which means that the jets dual to these endpoints will be quenched quickly and hence won't be observable. Therefore, for observable, partially quenched jets, the strings need to start rather close to the horizon\footnote{From the bulk perspective, one can consider collisions of shock waves in $AdS$ that have finite transverse extent and are sourced by some distribution of matter. The horizon forms around that matter and thus the location of rare energetic string formation is more naturally near the horizon and not in the middle of the bulk.}. For $z_*<z$ we see that the expansion \eno{eloss3} is strongly convergent and results in a particularly interesting novel form for energy loss:
\eqn{eloss4}
{\frac{dE}{dx}=-\frac{\pi}{2}\sqrt{\lambda}T^2\left(\frac{1}{\tilde z_0}+\pi T x\right)^2\,,}
where $\tilde z_0\equiv \pi T z_0 \in [0,1]$. This form of energy loss has an interesting physical interpretation: at small $x$, it looks like a pure $\sim T^2$ energy loss, similar to the pQCD elastic energy loss (with a running coupling); for intermediate $x$, it looks like $\sim x T^3$ with a path dependence (but not the energy dependence) similar to the pQCD radiative energy loss; and, finally, for large $x$, it has a novel $\sim x^2 T^4$ behavior\footnote{$T^4$ scaling of energy loss was also found in \cite{gp11} for the case of heavy quarks, but in a somewhat different context, and with different dependence on $\lambda$ and $x$.}. The size of $\tilde z_0$ (i.e. how much above the horizon the endpoint starts) dictates at what $x$ each of these regimes becomes relevant. This is an interesting (and a very specific) generalization of the simpler ``abc'' models of energy loss \cite{hg11}, where $dE/dx\propto E^a x^b T^c$. Another feature worth pointing out is that this is an energy-independent energy loss, which is a direct and rather general consequence of the energy loss from finite momentum endpoints. Because of this, in the high-$p_T$ regime, when the production spectra assume a power-law form, we expect to obtain an $R_{AA}$ which rises with $p_T$.


\section{Conformal $R_{AA}$}
\label{confraa}

In this section we will use the proposed formula \eno{eloss4} for the energy loss to compute $R_{AA}$ for pions at RHIC and LHC\footnote{For one of the first computations of (heavy quark) $R_{AA}$ in the AdS/CFT context, see \cite{ngt10}.}. There are several steps that will be taken in order to compute a more realistic $R_{AA}$, the purpose of which is to imitate some of the features of QCD:
\begin{itemize}
\item The first step is to express all the energy variables in GeV's and length variables in femtometers.
\item To account for roughly three times more degrees of freedom in $\mathcal{N}=4$ SYM than in QCD, we will relate the temperatures via \cite{g07}:
\eqn{raa1}
{T_{\rm SYM}=3^{-1/4}\,T_{\rm QCD}\,.}
\item The next step is to promote a constant $T_{\rm QCD}$ to a Glauber-like $T_{\rm QCD}(\vec x_\perp,t,\phi)$.
\item We will introduce the transverse expansion via a simple blast wave dilation factor \cite{bg13}:
\eqn{raa3}
{r_{\rm bl}(t)=\sqrt{1+\left(\frac{v_T t}{R}\right)^2}\,,}
where $R$ is the mean nuclear radius. We will take the transverse velocity $v_T=0.6$. The effect of this dilation factor will be to replace $\rho_{\rm part}(\vec x_\perp)\to \rho_{\rm part}(\vec x_\perp/r_{\rm bl})/r_{\rm bl}^2$ in the Glauber model.
\item Finally, we will use the fragmentation functions \cite{kkp00}\footnote{Another more natural feature of shooting strings that differs from the falling strings is that the virtuality of the endpoint, $Q^2\equiv p_0^2-p_x^2$, is proportional to the endpoint's energy squared, $Q^2=E^2\tilde z_*^4/(1-\tilde z_*^4)$. This energy, and hence virtuality as well, decreases even during the ``ascending'' phase in the geodesic trajectory, in a unique way given by \eno{eloss1} and \eno{eloss2}.  However, at finite $N_c$, one should bear in mind that even bulk constructions can be off-shell; thus some $N_c$ suppressed contribution to virtuality may be significant compared to the rather suppressed classical expression just given. For this reason, we will use the usual prescription of $Q=p_T$, and note that, due to a, for our purposes, low sensitivity of the fragmentation functions to the $Q^2$-evolution, we do not expect that a choice of a different prescription would affect our results significantly.} to obtain the pionic $R_{AA}$ from the partonic one (neglecting the gluon contribution\footnote{Although at the pure partonic level, gluons dominate the spectrum up to very high $p_T$, after quenching (gluon quenching being enhanced by the 9/4 coming from the ratio of Casimirs) and then fragmentation to pions (which is softer, with $p_\pi\approx 0.5 p_g$), $R_{AA}$ is dominated by the quark jets. See e.g. Fig. 27 in \cite{JiechenCUJET2} and Fig. 3 in \cite{hg11}. We note that an enhancement factor of 2 (close to the 9/4 factor from the ratio of Casimirs) in an AdS/CFT model of gluon energy loss was proposed in \cite{ggpr08}.}). 
\end{itemize}

We will use the standard optical Glauber model\footnote{We are aware that a more realistic treatment of fluctuating initial geometry and transverse flow will be needed in future for more quantitative applications of Eq. \eno{eloss4} to A+A phenomenology, but the optical model used here is sufficient to demonstrate that Eq. \eno{eloss4} comes closer to the data than previous holographic proposals.} to compute the participant and binary collisions densities, include the effects of longitudinal expansion and model the spacetime evolution of the temperature.  If a jet is created at position $\vec x_\perp$ in the transverse plane at time $t=0$ and moves radially at angle $\phi$, then the local temperature that it sees at some later time $t$ is given by
\eqn{raaextra}
{T(\vec x_\perp,t,\phi)=\left[\frac{\pi^2}{\zeta(3)}\frac{1}{16+9n_f}\frac{dN/dy}{N_{\rm part}}\frac{\rho_{\rm part}\left(\vec x_\perp +t\hat e(\phi)\right)}{t+t_i}\right]^{1/3}\,,}
where $n_f$ is the number of active flavors (which we will take to be 3), $dN/dy$ is the multiplicity and $t_i$ is the plasma equilibration (formation) time (which will be typically between 0.5 and 1 fm/c). For this jet we can find, using the energy loss formula \eno{eloss4}, its initial energy $p_{T,i}$, provided it has a fixed final energy $p_{T,f}$ at the time when the temperature reaches the freezout temperature $T_{\rm freeze}$. Averaging the ratio of the initial production spectra $d\sigma/dp_T$ (obtained from the LO pQCD CTEQ5 code \cite{WangPrivate}) at final and initial energies over the transverse plane we get the nuclear modification factor:
\eqn{raa5}
{\left<R_{AA}^\phi\right>(p_{T,f})=\int d^2\vec{x}_\perp \,\frac{T_{AA}(\vec{x}_\perp)}{N_{bin}}\frac{d\sigma/dp_T(p_{T,i}(p_{T,f}))}{d\sigma/dp_T(p_{T,f})}\,,}
where $T_{AA}$ is the number density of binary collisions and $N_{bin}$ is the total number of binary collisions.  For $\phi=0$ we obtain $R_{AA}^{\rm in}$ from \eno{raa5}, while for $\phi=\pi/2$ we get $R_{AA}^{\rm out}$. For nucleon-nucleon inelastic cross sections we will use $\sigma_{NN}^{RHIC}=42$ mb and $\sigma_{NN}^{LHC}=63$ mb \cite{dwx11}. We will also use the mixed participant-binary number scaling of the multiplicities in the non-central collisions with 85\% of participant scaling and 15\% of binary number scaling. For charged multiplicities we use $dN^{RHIC,ch}/d\eta=700$ and $dN^{LHC,ch}/d\eta=1584$. 

\begin{figure*}
\centerline{\includegraphics[width=3.35in]{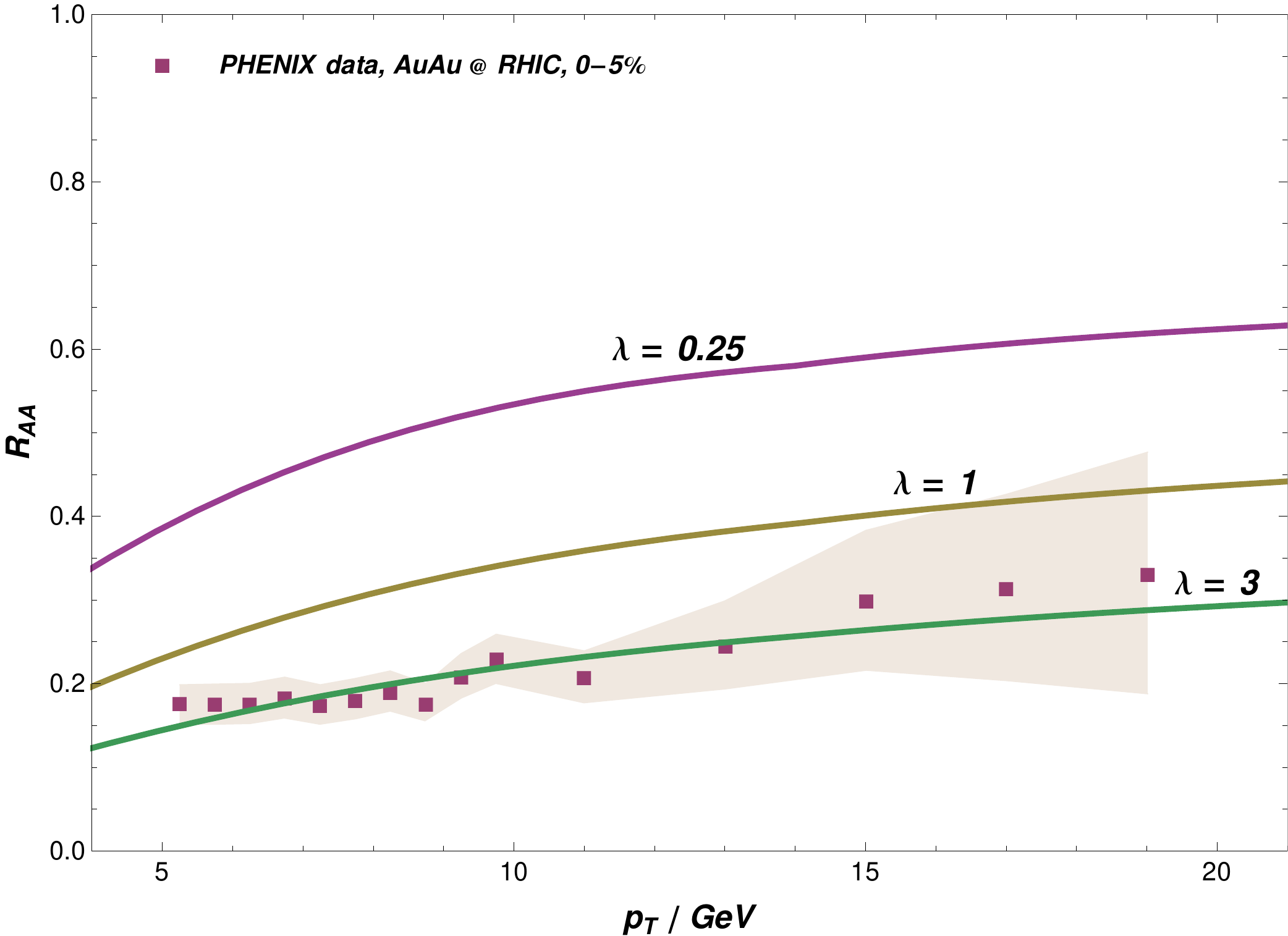}\hskip0.3in
\includegraphics[width=3.35in]{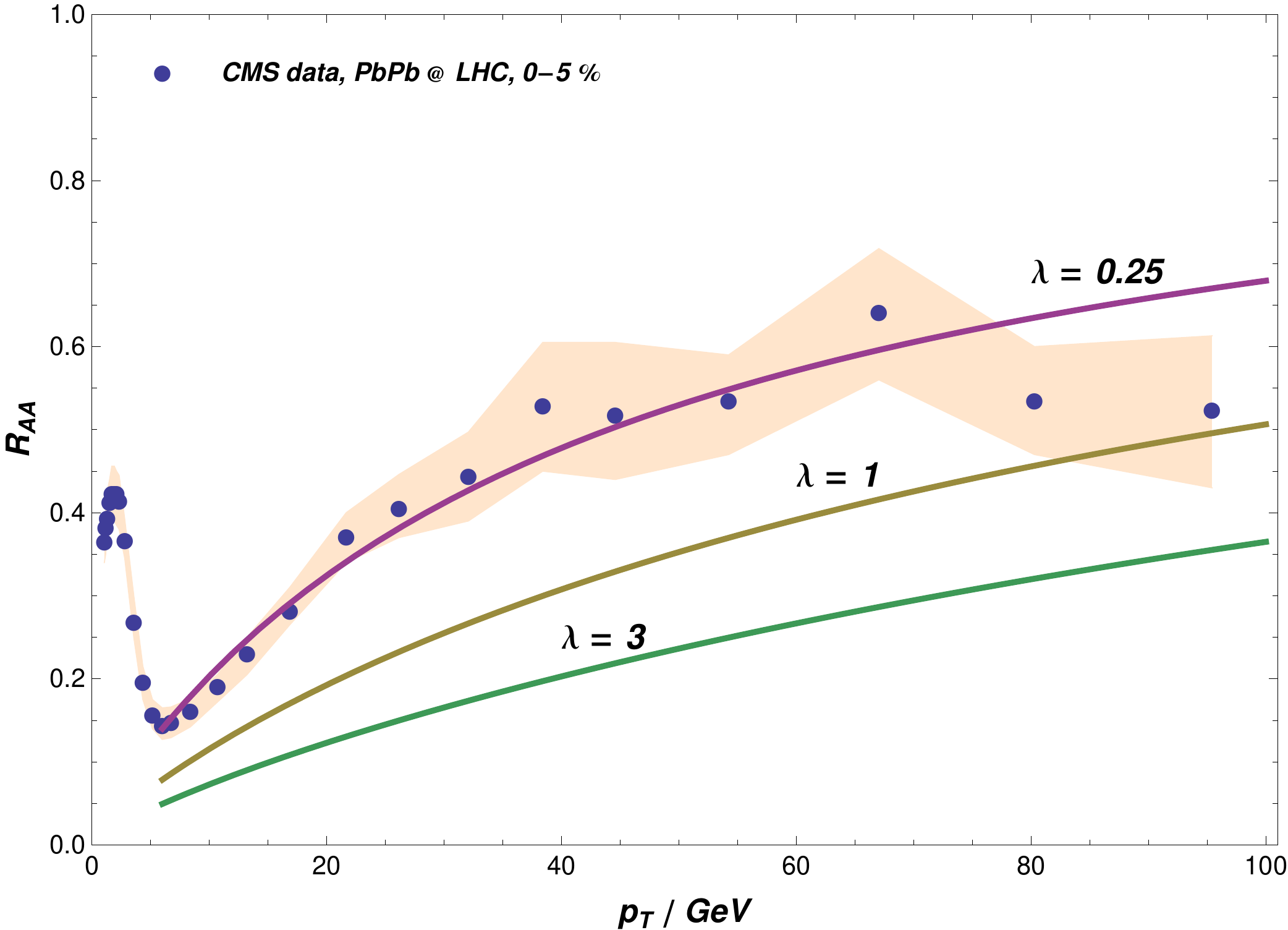}}
\caption{Nuclear modification factor $R_{AA}$ of pions in central collisions at RHIC and LHC. Our calculations are compared to the experimental data from the PHENIX \cite{phenix12} and the CMS \cite{cms12} collaborations for 0-5 \% centrality class. In different plots we only change the 't Hooft coupling $\lambda$ while they all have the same impact factor of $b=3$ fm, the freezout temperature of $T_{\rm freeze}=170$ MeV, the formation time of $t_i=1$ fm/c and the initial $\tilde z_0=1$ (from \eno{eloss4}).}
\label{RHICLHCCentral}
\end{figure*}

The results are shown in Fig. \ref{RHICLHCCentral}, where we chose (as in the rest of the plots in the paper) $\tilde z_0=1$. There we see that, first of all, qualitatively, as expected, our $R_{AA}$ calculations seem to match the data well. To obtain a satisfactory quantitative fit, with a reasonable choice of parameters, we needed to choose $\lambda=3$ at RHIC; however, using the same parameters and $\lambda$ at the LHC shows that the data is severely underpredicted. Lowering $\lambda$ to 1 for LHC data is not enough: one seems to need a radically small $\lambda=0.25$ to obtain a satisfactory fit. Of course, with that $\lambda$, RHIC is then severely overpredicted. This is precisely the ``surprising transparency'' of the LHC \cite{hg11}, where the effects of temperature increase from RHIC to LHC affect the $R_{AA}$ much more than the competing increase of the production spectra \footnote{This may be partially resolved by considering duals of non-conformal field theories where the coupling gets an effective temperature running. For example, by introducing a specific potential for the dilaton, one can construct a bottom-up non-conformal deformation of $\mathcal{N}=4$ SYM \cite{gnpr08,fng11} that has the same thermodynamics (and Polyakov loops) as the one provided by lattice QCD. In those models the running of the dilaton causes the energy loss at low temperatures to increase relative to the conformal limit, which in turn affects the $R_{AA}$ at RHIC more than at the LHC, but this effect was not strong enough to resolve the problem entirely.}. 

\section{Higher derivative corrections}
\label{gb}

A possible way to make our setup more realistic is to add higher derivative $R^2$-corrections to the gravity sector of $AdS_5$, which are the leading $1/N_c$ corrections in the presence of a $D7$-brane. It has been shown \cite{fng12} that these type of corrections can increase $R_{AA}$ significantly and in this section we will explore their effect in the context of finite endpoint momentum strings\footnote{The effects of the $R^2$-corrections on the drag force of a heavy quark (represented by the usual trailing string with no endpoint momentum) were investigated in \cite{f08}.}.

We will model the $R^2$ corrections by a Gauss-Bonnet term, i.e. we will consider the action of the form: 
\eqn{gb1}{
\begin{aligned}
S&=\frac{1}{2\kappa_5^2}\int d^5 x\sqrt{-G}\Bigg[R+\frac{12}{L^2}  \cr
  &\qquad{} +L^2\frac{\lambda_{GB}}{2}\left(R^2-4R_{\mu\nu}^2+R_{\mu\nu\rho\sigma}^2\right)\Bigg]\,,
  \end{aligned}}
where $\lambda_{GB}$ is a dimensionless parameter, constrained by causality \cite{blmsy08} and positive-definiteness of the boundary energy density \cite{hm08} to be:
\eqn{gb2}
{-\frac{7}{36}<\lambda_{GB}\le \frac{9}{100}\,.} 
A black hole solution in this case is known analytically \cite{c02}:
\eqn{gb3}
{ds^2=\frac{L^2}{z^2}\left(-a^2f_{GB}(z)dt^2+dx^2+\frac{dz^2}{f_{GB}(z)} \right)\,,}
where
\eqn{gb4}
{\begin{aligned}
f_{GB}(z)&=\frac{1}{2\lambda_{GB}}\left(1-\sqrt{1-4\lambda_{GB}(1-z^4/z_H^4)}\right)\,,\cr
a^2&=\frac{1}{2}\left(1+\sqrt{1-4\lambda_{GB}}\right)\,.
\end{aligned}}
The 't Hooft coupling and the temperature are given by
\eqn{gb5}
{\sqrt{\lambda}=a^2\frac{L^2}{\alpha'},\qquad T=\frac{a}{\pi z_H}\,.}

Using the same procedure as before (reviewed in Appendix \ref{GeneralShoot}), we can easily find the energy loss from the finite endpoint momentum in geometry \eno{gb3} and obtain a formula similar to \eno{eloss1}:
\eqn{gb6}
{\frac{dE_{GB}}{dx}=- \frac{\sqrt{\lambda}}{2\pi}\frac{1}{z^2}\frac{\sqrt{f_{GB}(z_*)}}{a}\,.}
One can obtain this expression immediately from \eno{eloss1} by noting that $AdS_5$-Schwarzschild differs from \eno{gb3} by replacing $f(z)$ with $f_{GB}(z)$ and $E$ with $E/a$. To express \eno{gb6} this as a function of $x$, we need to solve for null geodesics:
\eqn{gb7}
{\frac{dx_{\rm geo}}{dz}=\pm \frac{1}{\sqrt{f_{GB}(z_*)-f_{GB}(z)}}\,.}
To obtain the generalization of the formulas from the previous section, we will send $z_*\to 0$ here and then, for a given $\lambda_{GB}$, we can easily numerically integrate \eno{gb7} and invert to obtain $z(x)$, which can then be plugged in \eno{gb6} to get $dE/dx$ as a function of $x$ and $T$. 

However, since according to \eno{gb2}, $\lambda_{GB}$ is constrained to be small, these expressions are suitable for a perturbative expansion in $\lambda_{GB}$, allowing for a more practical analytic expression. To do this, we will expand \eno{gb7} in $\lambda_{GB}$ up to some order $n$ and neglect all terms higher than $1/z^2$, as they are $\mathcal{O}(z^4)$ subleading. Of course, we will be able to check how accurate this is by comparing with the full numerical solution. We define a polynomial in $\lambda_{GB}$:
\eqn{gb8}
{P_n(\lambda_{GB})\equiv \frac{2}{z_H^2}\lim_{z\to0} z^2\left(\frac{dx_{\rm geo,(n)}}{dz}\right)_{z_*=0}\,,}
where $n$ denotes the order of expansion in $\lambda_{GB}$. In this case, we can easily solve the geodesic equation:
\eqn{gb9}
{z_n(x,\lambda_{GB})=\frac{z_H^2 z_0 P_n(\lambda_{GB})}{z_H^2 P_n(\lambda_{GB})-2x z_0}\,.}
This can be plugged in \eno{gb6} to get explicitly the form of $dE/dx$ for a given order $n$, yielding an expression similar to \eno{eloss4}:
\eqn{gb10}
{\frac{dE_{GB}}{dx}=-\sqrt{\lambda}\,T^2 F_n(\lambda_{GB})\left(\frac{G_n(\lambda_{GB})}{\tilde z_0}+\pi T x\right)^2\,.}
The functions $F_n$ and $G_n$ are functions of $\lambda_{GB}$ only and do not have a particularly illuminating explicit form, even for small $n$. For $\lambda_{GB}$ as large as $-7/36$, by comparing to the all-order numerical result, we found that it is enough to go to $n=5$ order in expansion.

\begin{figure*}
\centerline{\includegraphics[width=3.35in]{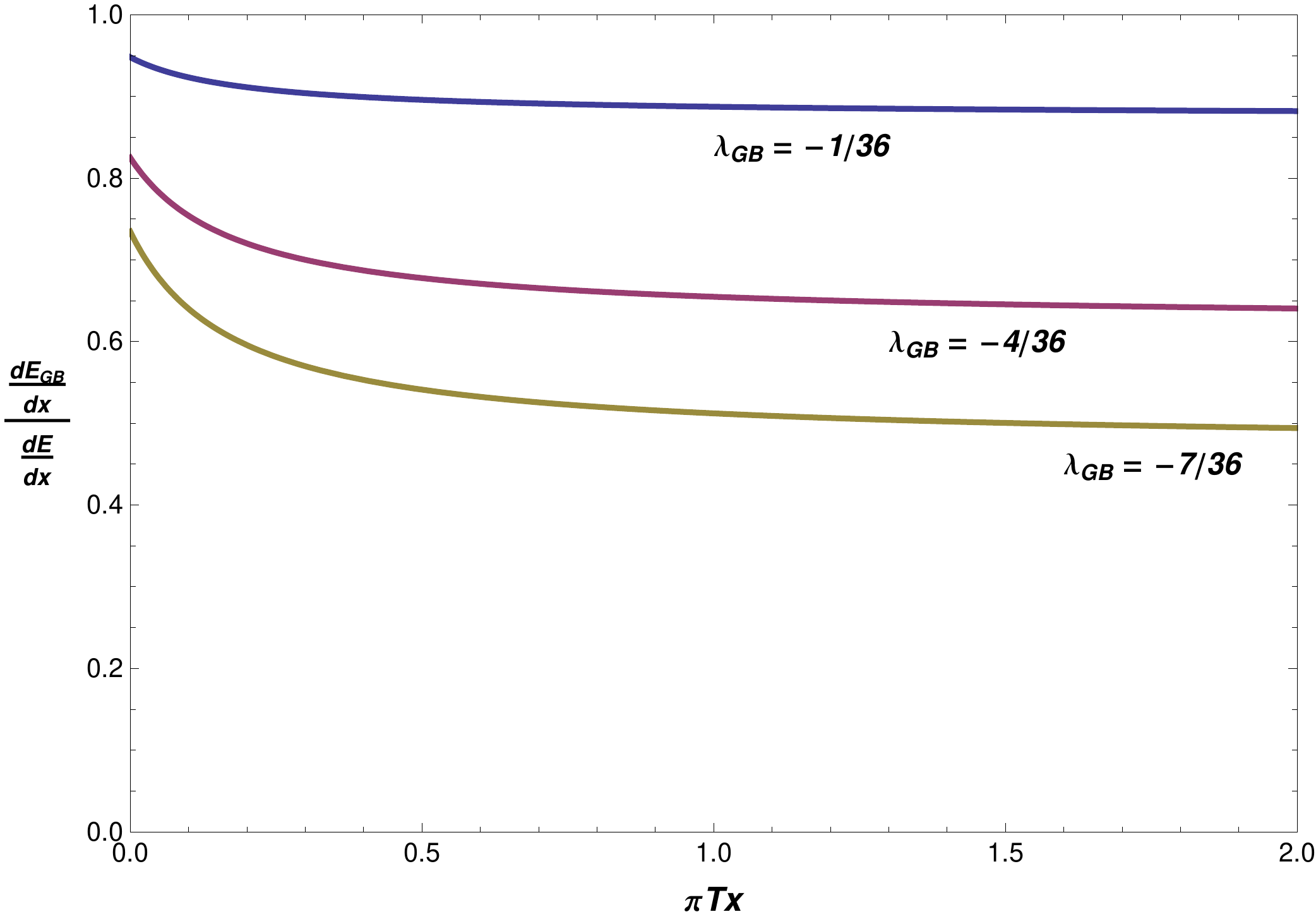}
\includegraphics[width=3.35in]{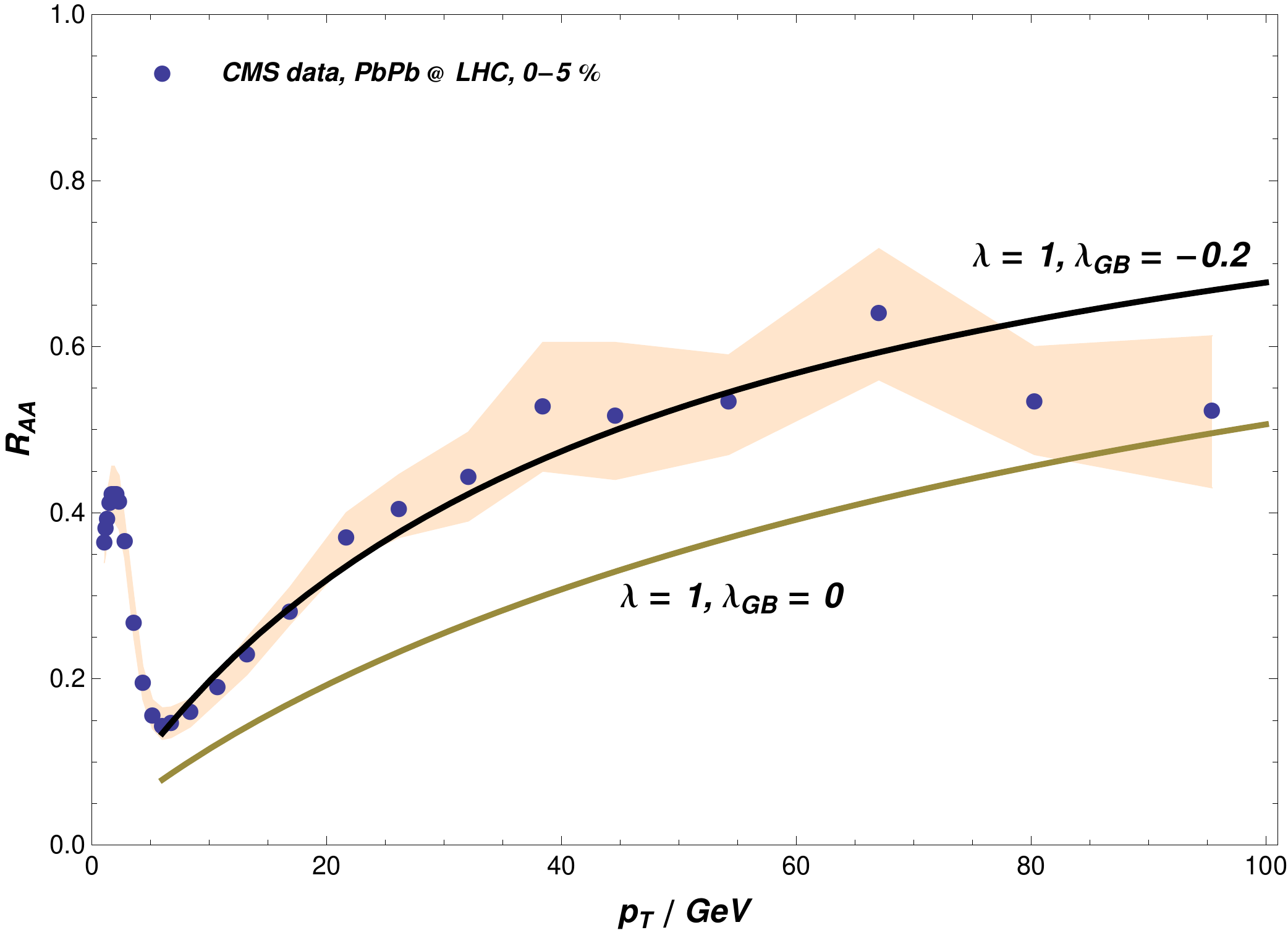}}
\caption{{\it Left:} Ratio of the instantaneous energy loss in pure $AdS$ \eno{eloss4} and the energy loss with the Gauss-Bonnet corrections \eno{gb10} as a function of $x$, for $\tilde z_0=1$ and for several different values of $\lambda_{GB}$. {\it Right:} Nuclear suppression factor $R_{AA}$ at the LHC for $\lambda=1$, with and without the higher derivative corrections. All other parameters are the same as in Fig. \ref{RHICLHCCentral}.}
\label{GBvsCFT}
\end{figure*}

In the left plot of Fig. \ref{GBvsCFT} we compare this energy loss to the energy loss without the Gauss-Bonnet term, where we can see that, at a maximally negative $\lambda_{GB}$, the energy loss with the Gauss-Bonnet corrections can be up to two times smaller. This, expectedly, has noticeable consequences for the $R_{AA}$ (right plot of Fig. \ref{GBvsCFT}): we see that it results in a higher $R_{AA}$ that comes very close to the data for $\lambda=1$. Recalling that in the presence of the Gauss-Bonnet term, the shear viscosity is given by \cite{blmsy08}:
\eqn{gb11}
{\frac{\eta}{s}=\frac{1-4\lambda_{GB}}{4\pi}\,,} 
(which is an exact result, up to all orders in $\lambda_{GB}$), we see that negative values of $\lambda_{GB}$ increase the viscosity. For a maximally negative $\lambda_{GB}$, the viscosity can be increased up to about $1.8/(4\pi)$, which is, together with our selected value of the initial time $t_i=1$ fm/c, in the ballpark of the parameters used in the most recent hydrodynamic simulations for the LHC \cite{sbh13} to describe the elliptic flow data of light hadrons.

\begin{figure*}
\centerline{\includegraphics[width=3.35in]{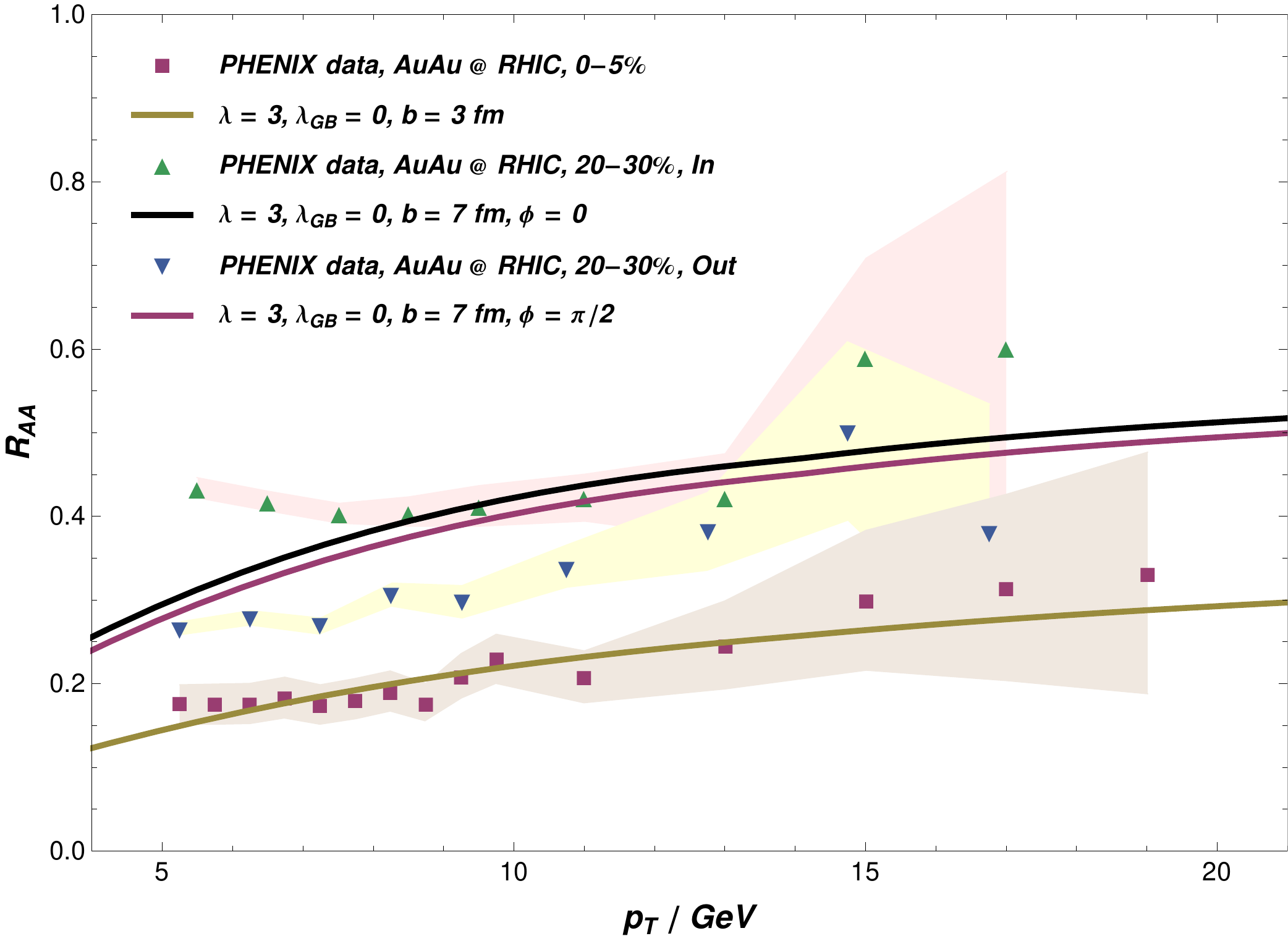}\hskip0.3in
\includegraphics[width=3.35in]{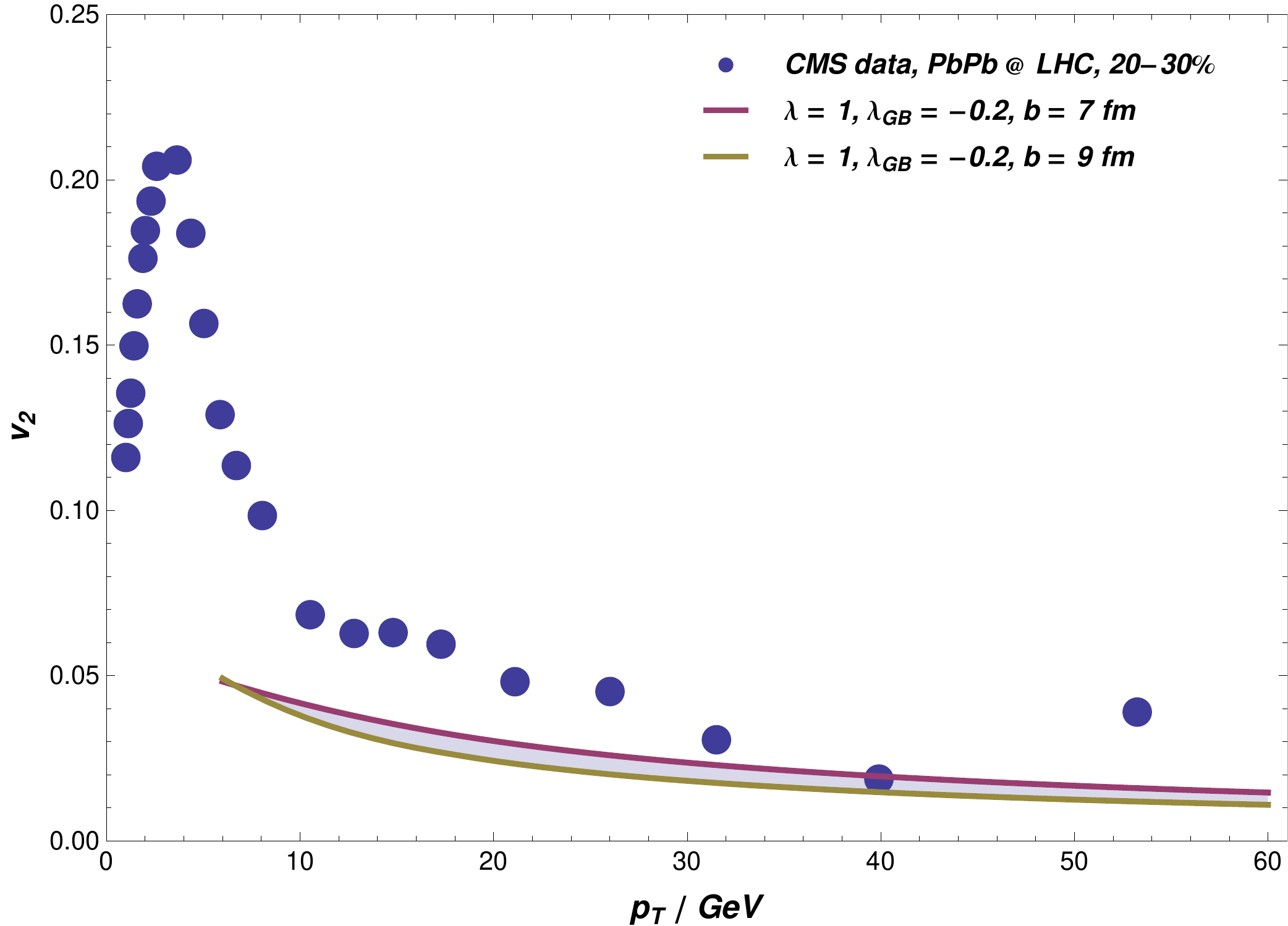}}
\caption{Nuclear modification factor at RHIC in non-central collisions (left plot) and the elliptic flow parameter at the LHC (right plot). The experimental data for both RHIC \cite{phenix12} and LHC \cite{cms12a} are for the 20-30 \% centrality class. In the left plot we compare the $R_{AA}$ in central collisions as well as the {\it in} and {\it out} $R_{AA}$ in non-central collisions to our calculations for $b=3$ fm and $b=7$ fm, respectively. In the right plot, the band corresponds to the $v_2$ calculations for $b$ between 7 and 9 fm. All the other parameters in these plots are the same as in Fig. \ref{RHICLHCCentral}.}
\label{RHICLHCNoncentral}
\end{figure*}

Now that we have the central $R_{AA}$ data well matched for a reasonable choice of parameters both at RHIC and LHC, we can inspect what happens in the non-central case. In that case, we will compute the elliptic flow parameter using the approximate formula \cite{JiechenCUJET2}:
\eqn{gb12}
{v_2\approx\frac{1}{2}\frac{R_{AA}^{\rm in}-R_{AA}^{\rm out}}{R_{AA}^{\rm in}+R_{AA}^{\rm out}}\,.}
In Fig. \ref{RHICLHCNoncentral} we see that in the case of RHIC, the splitting we predict for {\it in} and {\it out} $R_{AA}$'s in non-central collisions is not big enough, which is probably due to the (too) simple blast wave we are using to model the transverse expansion of the plasma. We see a similar result in the case of LHC as well, where a too small {\it in-out} splitting results in $v_2$ that seems to be somewhat below the data for the $\lambda=1$ case with the Gauss-Bonnet corrections, which matched the central $R_{AA}$ data.

\section{RHIC vs. LHC and the temperature sensitivity}
\label{temp}

It is clear that the choice of $\lambda=1$, including the higher derivative corrections with $\lambda_{GB}=-0.2$ (which give $\eta/s=1.8/(4\pi)$), that matches the LHC central $R_{AA}$ data (black curve in Fig. \ref{GBvsCFT}) will result in a significant overprediction of the central RHIC data; $\lambda=1$ with $\lambda_{GB}=-0.2$ at RHIC approximately corresponds to the $\lambda=0.25$  case with $\lambda_{GB}=0$, i.e. the purple curve in Fig. \ref{RHICLHCCentral}. Hence the simultaneous fit of the RHIC and the LHC central $R_{AA}$ data remains a challenge in our simple constructions presented here.  But we would like to point to a possible phenomenological effect that can be partially responsible for this discrepancy: the effective temperature uncertainties. Note that our energy loss formula \eno{eloss4} has a strong sensitivity to the temperature, $dE/dx\sim \sqrt{\lambda} T^3$ or even $\sqrt{\lambda}T^4$. Hence, even a small change in the temperature, $T\to \kappa T$ can have the same effect as a large change in the coupling, $\lambda\to\kappa^6\lambda$ or $\kappa^8\lambda$. 

\begin{figure*}
\centerline{\includegraphics[width=3.35in]{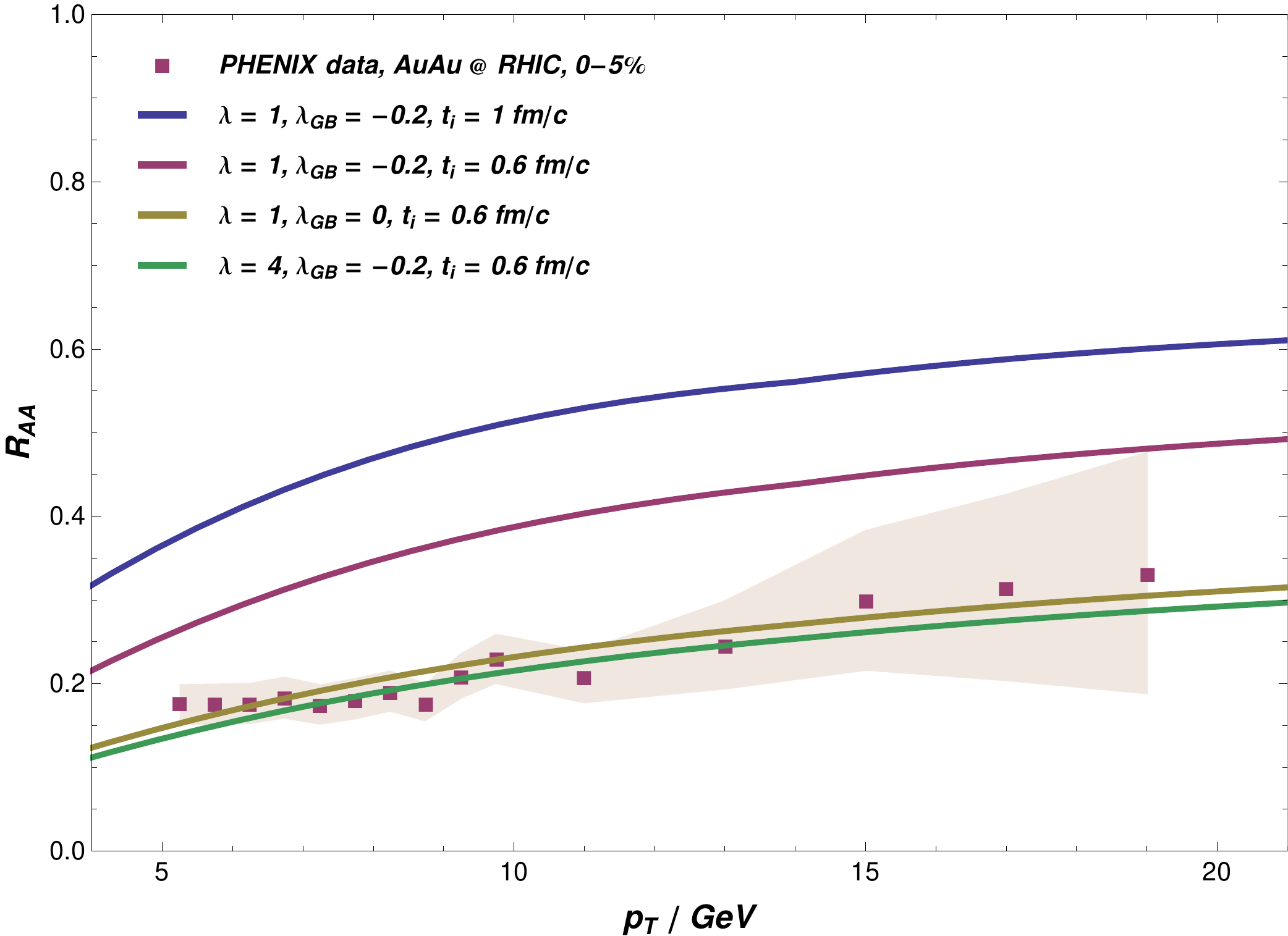}\hskip0.3in
\includegraphics[width=3.35in]{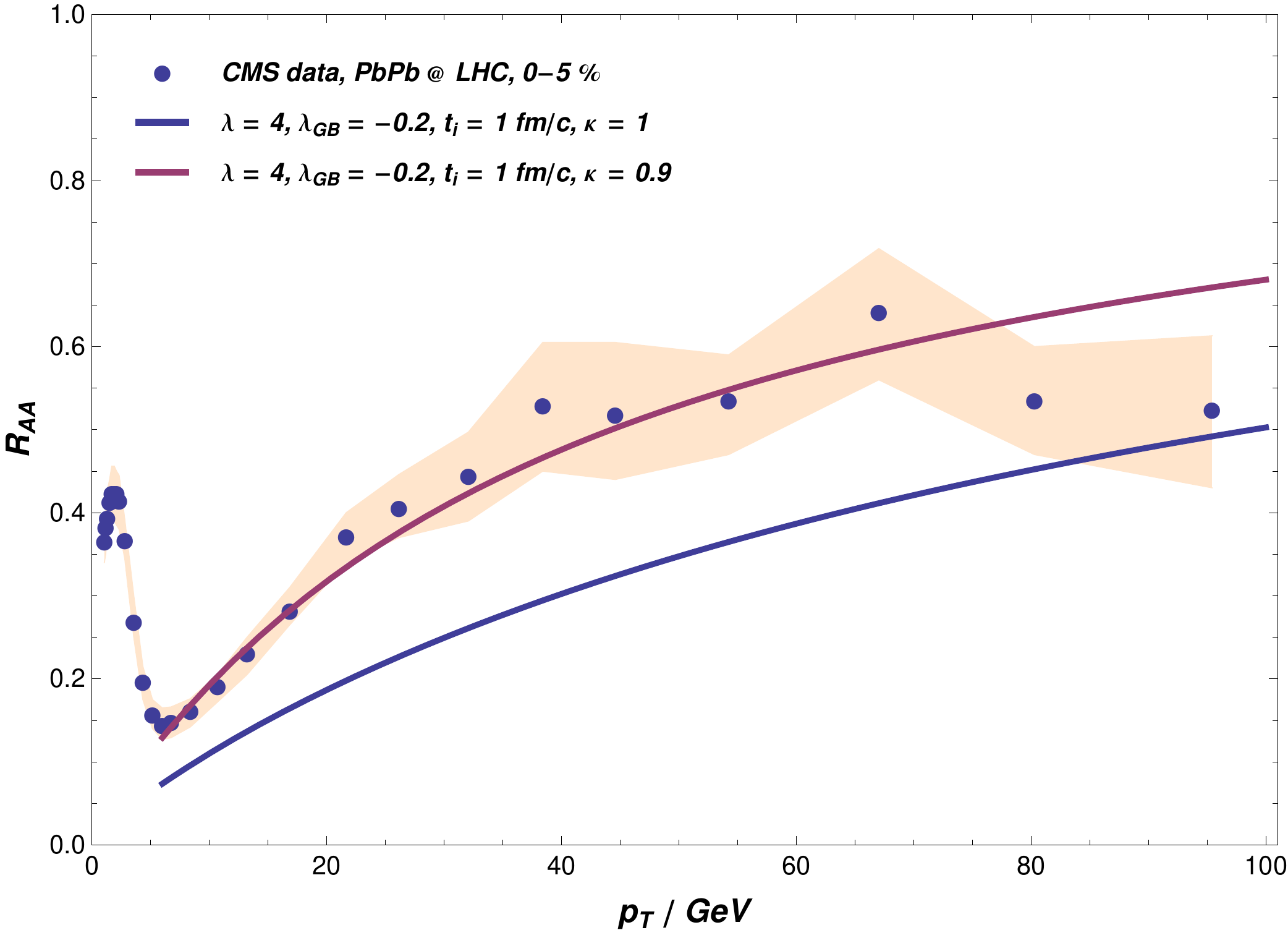}}
\caption{{\it Left:} Nuclear modification factor at RHIC in central collisions for different choices of the 't Hooft coupling $\lambda$, dimensionless $\lambda_{GB}$ and the initial time $t_i$. {\it Right:} Nuclear modification factor at the LHC in central collisions with and without the temperature adjustment $T\to \kappa T$.}
\label{TempPlots}
\end{figure*}

We cannot offer at the moment a concrete physical reason that would justify the possibility of temperature uncertainties, but we can speculate based on some very general arguments. From the perspective of the temperature formula \eno{raaextra} we see that one would expect the LHC to be roughly 30\% hotter than RHIC, based on the ratio of the multiplicities. However, if the initial time $t_i$ in the two cases is different, then the jet effectively feels a cooler or a hotter medium, according to \eno{raaextra}. This is precisely what was suggested in \cite{sbh13}, where the authors used a bigger initial time at the LHC than at RHIC where $t_i=0.6$ fm/c \cite{sbhhs11}, based on the requirements of the hydrodynamic simulations to fit the low $p_T$ elliptic flow data. We see the effect of this in the left plot of Fig. \ref{TempPlots}, where changing $t_i$ from 1 fm/c (blue) to 0.6 fm/c (red) for the case of $\lambda=1$ and $\lambda_{GB}=-0.2$ (which fits the central LHC data) leads to a significant decrease in $R_{AA}$. Additionally, if we allow $\eta/s$ to decrease (relative to LHC, where the temperature range is higher), which means increasing $\lambda_{GB}$ (as per \eno{gb11}), we can approach the RHIC data even more (yellow curve). We should note that this is just an illustration of the effect of the decrease of $\eta/s$ on $R_{AA}$, as the same hydrodynamic calculations of \cite{sbhhs11} and \cite{sbh13} suggest that this decrease is not so strong. If we want to keep the same $\eta/s$ at RHIC as it was at the LHC (meaning keeping $\lambda_{GB}=-0.2$) then we can get close to the data by increasing the coupling approximately 4 times (green curve). If we keep these parameters of the green curve and pass onto LHC (right plot of Fig (\ref{TempPlots})), where we set $t_i=1\,{\rm fm/c}$, we see that the curve is below the data (blue curve), but lowering the overall LHC temperature by about 10\% we are able to approach the data (red curve).

\section{Conclusions}
\label{conclusions}

In this Letter we have proposed a novel formula \eno{eloss4} for the instantaneous energy loss of light quarks in a strongly coupled SYM plasma. This formula was derived in the framework of finite endpoint momentum strings, where the jet energy loss is identified with the energy flux from the endpoint to the bulk of the string, offering a clear definition of the energy loss that is independent of the details of the string configuration and results in greater stopping distances. Application of this formula, using endpoints that start close to the horizon (``shooting'' strings) and including the higher derivative $R^2$-corrections showed, independently, a very good match with the RHIC and LHC central $R_{AA}$ data for light hadrons, and even partially for the elliptic flow $v_2$. A consistent simultaneous match of both the RHIC and the LHC central $R_{AA}$ data remains challenging, but, as we argue, the temperature sensitivity of our formula coupled with the uncertainties in the formation time $t_i$ and the shear viscosity $\eta/s$ at RHIC and LHC may enable us to reconcile these differences. In particular, we have shown that, using a smaller formation time at RHIC and perhaps allowing for a slightly smaller $\eta/s$ than at the LHC, one can significantly reduce the RHIC-LHC splitting. Further inclusion of non-conformal effects, which are known to moderately increase the energy loss at lower temperatures (and hence affect RHIC more than the LHC), may provide an additional reduction of the splitting.


\section*{Acknowledgments}

The work of A.F.\ and M.G.\ was supported by U.S. DOE Nuclear Science Grant No. DE-FG02-93ER40764.  The work of S.S.G.\ was supported in part by the Department of Energy under Grant No.~DE-FG02-91ER40671.


\appendix
\section{Energy loss from falling strings revisited}
\label{fallGB}

In this section we will use the methods of \cite{fg13} to provide additional support and better justification of the main assumptions behind the phenomenological model of energy loss \cite{fng12} based on falling strings, that was used in obtaining the dashed curves in Fig. \ref{IntroFig}.

\subsection{Linear path dependence of energy loss}
The first result used in the model of \cite{fng12} was the linear path dependence of the energy loss, $dE/dx\propto x$, in the phenomenologically relevant range, that was first suggested in \cite{f12} through preliminary numerical studies. Using the methods of \cite{fg13}, we will provide additional analytical support to this numerical observation. The main point of \cite{fg13} was to recognize that the energy of an energetic falling string can be well approximated by the UV part of the energy of a trailing string \cite{g06,hkkky06} without the drag force term, whose endpoint is moving at the same radial height at the local speed of light:
\eqn{tt1}{
E_*=\frac{\sqrt{\lambda}}{2\pi}\frac{1}{z_*}\frac{1}{\sqrt{1-v^2}}\,,
}
for a falling string in $AdS_5$-Schwarzschild whose endpoint is at $z=z_*$ and is moving at the local speed of light $v$ in the $x-$direction. Using this result, together with the fact that the endpoints approximately follow null geodesics whose minimum radial distance to the boundary is $z_*$, resulted in an analytical expression \cite{fg13} for the stopping distance of light quarks, confirming the numerical results of \cite{cjky08}. The endpoints stay close to $z_*\ll z_H$ for a long time compared to $z_*$ and from the null geodesic equation we can easily get that in this regime
\eqn{tt2}{
\frac{dz_{\rm geo}}{dx}=\frac{2z_*^3}{z_H^4}x\left[1+\mathcal{O}\left(\frac{x^2z_*^2}{z_H^4}\right)\right],\
}
where, for simplicity, we assumed that at $x=0$ the endpoint was at $z=z_*$. To get the energy loss, we can simply take the derivative of \eno{tt1} where now $z$ and $v$ are slowly changing as the endpoint is falling down:
\eqn{tt3}{
\frac{dE}{dx}=\frac{\sqrt{\lambda}}{2\pi}\frac{d}{dz}\left(\frac{1}{z}\frac{1}{\sqrt{1-v(z)^2}}\right)\frac{dz_{\rm geo}}{dx}\,.
}
Again, we are interested in the regime where the endpoints stay close to $z_*$, which lasts arbitrarily long in the small $z_*$ (high energy) limit. In this case, the leading, $z-$independent term in the expansion in $(z-z_*)$ of the term in front of $dz_{\rm geo}/dx$ is finite and non-zero, which means that the only $x-$dependence in \eno{tt3} comes from \eno{tt2}, hence showing that the energy loss is linear in $x$ to the leading order in small $z_*$, and thus reaffirming the numerical indications of \cite{f12}.

\subsection{Stopping distance in $AdS_5$ with Gauss-Bonnet corrections}
The second result used in the constructions of \cite{fng12} is the stopping distance for a falling string in $AdS_5$ geometry with the Gauss-Bonnet corrections \eno{gb1}. We will again use the analytical procedure from \cite{fg13}, with the aim to provide a more clear and reliable derivation of this, rather than using the preliminary result reported in \cite{fng12}. Again, the idea is to compute the energy of a trailing string hanging from $z_*$, discard the IR-divergent drag force term and take the UV limit $z_*\ll z_H$. Then we compute the range of the null geodesic that starts at that $z_*$ parallel to the boundary and again take the $z_*\ll z_H$ limit. All this is done perturbatively in $\lambda_{GB}$, keeping only the linear terms (although it is straightforward to go to higher orders). Finally, this range and the ``regularized'' energy are related via $z_*$ that folds into the final answer \eno{e7}.

If we define $\tau\equiv at$, the $AdS_5$-GB geometry \eno{gb3} has the same form as $AdS_5$-Schwarzschild with $f(z)$ replaced by $f_{GB}(z)$ and we can easily follow the derivation for the energy of the trailing string there (as e.g. in \cite{hkkky06}), keeping $f_{GB}(z)$ general. After subtracting the drag force term and multiplying by $a$ to get the energy conjugated to $t$, we arrive at:
\eqn{e2}
{E=\frac{L^2}{2\pi\alpha'}\frac{a}{z_*^2}\int\limits^{z_H}_{z_*}dz\frac{1}{z^2}\left[\frac{z^4 f_{GB}(z_*)-z_*^4 f_{GB}(z)}{f_{GB}(z_*)-f_{GB}(z)}\right]^{1/2}\,.}
Note that in the case of $AdS_5$-Schwarzschild, the term in the brackets is equal to $z_H^4$, yielding a simple $\propto 1/z^2$ integrand. 

We can plug in the expressions \eno{gb4} and \eno{gb5} in $E\equiv \int \varepsilon dz $ and the null geodesic, and expand in $\lambda_{GB}$:
\begin{widetext}
\begin{eqnarray}\label{e3}
\varepsilon&=&\frac{\sqrt{\lambda}}{2\pi^3 T^2 z^2 z_*^2}\left[1-\left(1+\frac{1}{2}\pi^8 T^8 z^4 z_*^4-\frac{1}{2}\pi^4 T^4 \left(z^4+z_*^4\right)\right)\lambda_{GB}+\mathcal{O}\left(\lambda_{GB}^2\right)\right]\,, \\ \label{e5}
\frac{dx_{\rm geo}}{dz}&=&\frac{1}{\sqrt{f_{GB}(z_*)-f_{GB}(z)}}=\frac{1}{\pi^2 T^2\sqrt{z^4-z_*^4}}\left[1-\left(2-\frac{1}{2}\pi^4 T^4\left(z^4+z_*^4\right)\right)\lambda_{GB}+\mathcal{O}\left(\lambda_{GB}^2\right)\right]\,.
\end{eqnarray}
\end{widetext}
Since we are interested in the $z_*\ll z_H$ limit, only the first term in the $\mathcal{O}(\lambda_{GB})$ order matters and we get:
\begin{eqnarray} \label{e4}
E&=&\frac{\sqrt{\lambda}}{2\pi^3 T^2}\frac{1}{z_*^3}(1-\lambda_{GB})\,,\\ \label{e6}
\Delta x&=&\frac{\Gamma\left(\frac{5}{4}\right)}{\pi^{3/2}T^2 \Gamma\left(\frac{3}{4}\right)}\frac{1}{z_*}\left(1-2\lambda_{GB}\right)\,.
\end{eqnarray}

We can now express $z_*$ in terms of $E$ from \eno{e4}, plug it in \eno{e6} and expand in $\lambda_{GB}$ yielding finally:
\eqn{e7}
{\Delta x=\left[\frac{2^{1/3}}{\sqrt{\pi}}\frac{\Gamma\left(\frac{5}{4}\right)}{\Gamma\left(\frac{3}{4}\right)}\right]\frac{1}{T}\left(\frac{E}{\sqrt{\lambda}T}\right)^{1/3}\left(1-\frac{5}{3}\lambda_{GB}\right)\,.}
The numerical term in front of $\lambda_{GB}$ quoted as a preliminary result in \cite{fng12} was $-11/6$, which was obtained by perturbative methods analogous to the ones in \cite{cjky08}. Here we have provided a transparent derivation of this result, with the obtained numerical factor being very close to the preliminary estimate of \cite{fng12}.


\section{Energy loss from finite momentum endpoints in more general geometries}
\label{GeneralShoot}

In this section we will provide a short derivation of the energy loss formula from a finite momentum endpoint, such as \eno{eloss1}, only in more general geometries. This should also serve as a demonstration of how simple it is to apply the finite endpoint momentum framework to arbitrary geometries, as well as to illustrate some general features of it.

We will assume that the spacetime metric has the following form:
\eqn{a1}{
ds^2=G_{tt}(z)dt^2+G_{xx}(z)dx^2+G_{zz}(z)dz^2\,.
}
The following simple derivation is easily applicable to metrics more general than this, but many cases of interest are captured by form \eno{a1}. Because the metric does not depend explicitly on $t$ nor $x$, the following is a constant of motion along a geodesic (following the notation of \cite{fg13}):
\eqn{a2}{
R=\frac{G_{tt}\dot t}{G_{xx}\dot x}\,,
}
where the dot denotes differentiation with respect to some parameter $\xi$ that parametrizes the geodesic. The finite momentum endpoints will move along null geodesics $ds^2=0$ parametrized by $R$:
\eqn{a3}{
\left(\frac{dx_{\rm geo}}{dz}\right)^2=-\frac{G_{tt}G_{zz}}{G_{xx}(G_{tt}+G_{xx}R^2)}\,.
}
If the geometry \eno{a1} allows for null geodesics such that the denominator of \eno{a3} vanishes at some $z=z_*$, then the geodesic cannot go past that (minimal) $z_*$ and we can relate it to $R$:
\eqn{a4}{
R=-\sqrt{\frac{-G_{tt}(z_*)}{G_{xx}(z_*)}}\,.
}

Because the metric \eno{a1} is not explicitly dependent on $t$, the flux of energy from the endpoint to the bulk of the string is given by a simple formula \cite{fg13}:
\eqn{a5}{
\dot p_t=- \frac{1}{2\pi \alpha'}G_{tt}\dot t\,.
}
This equation also explicitly demonstrates how the energy loss from a finite momentum endpoint does not depend on the energy contained in it, as the drain is caused by string worldsheet currents that do not know anything about the finite momentum except that it is there (as it changes the boundary conditions). Using $\xi=x$ parametrization and plugging \eno{a2} in \eno{a5} we get:
\eqn{a6}{
\frac{dE}{dx}=-\frac{|R|}{2\pi\alpha'}G_{xx}(z)\,.
}
In the case of $AdS_5$-Schwarzschild, we quickly arrive at \eno{eloss1}. Note that in the small $z_*$ limit, for asymptotically $AdS$ geometries, $R\to1$, but in order to find out how the energy loss depends on $x$ one must solve the null geodesic equation \eno{a3}.


\bibliography{pheno}

\begin{thebibliography}{29}%
\makeatletter
\providecommand \@ifxundefined [1]{%
 \@ifx{#1\undefined}
}%
\providecommand \@ifnum [1]{%
 \ifnum #1\expandafter \@firstoftwo
 \else \expandafter \@secondoftwo
 \fi
}%
\providecommand \@ifx [1]{%
 \ifx #1\expandafter \@firstoftwo
 \else \expandafter \@secondoftwo
 \fi
}%
\providecommand \natexlab [1]{#1}%
\providecommand \enquote  [1]{``#1''}%
\providecommand \bibnamefont  [1]{#1}%
\providecommand \bibfnamefont [1]{#1}%
\providecommand \citenamefont [1]{#1}%
\providecommand \href@noop [0]{\@secondoftwo}%
\providecommand \href [0]{\begingroup \@sanitize@url \@href}%
\providecommand \@href[1]{\@@startlink{#1}\@@href}%
\providecommand \@@href[1]{\endgroup#1\@@endlink}%
\providecommand \@sanitize@url [0]{\catcode `\\12\catcode `\$12\catcode
  `\&12\catcode `\#12\catcode `\^12\catcode `\_12\catcode `\%12\relax}%
\providecommand \@@startlink[1]{}%
\providecommand \@@endlink[0]{}%
\providecommand \url  [0]{\begingroup\@sanitize@url \@url }%
\providecommand \@url [1]{\endgroup\@href {#1}{\urlprefix }}%
\providecommand \urlprefix  [0]{URL }%
\providecommand \Eprint [0]{\href }%
\providecommand \doibase [0]{http://dx.doi.org/}%
\providecommand \selectlanguage [0]{\@gobble}%
\providecommand \bibinfo  [0]{\@secondoftwo}%
\providecommand \bibfield  [0]{\@secondoftwo}%
\providecommand \translation [1]{[#1]}%
\providecommand \BibitemOpen [0]{}%
\providecommand \bibitemStop [0]{}%
\providecommand \bibitemNoStop [0]{.\EOS\space}%
\providecommand \EOS [0]{\spacefactor3000\relax}%
\providecommand \BibitemShut  [1]{\csname bibitem#1\endcsname}%
\let\auto@bib@innerbib\@empty
\bibitem [{\citenamefont {Gubser}\ \emph
  {et~al.}(2008{\natexlab{a}})\citenamefont {Gubser}, \citenamefont {Gulotta},
  \citenamefont {Pufu},\ and\ \citenamefont {Rocha}}]{ggpr08}%
  \BibitemOpen
  \bibfield  {author} {\bibinfo {author} {\bibfnamefont {S.~S.}\ \bibnamefont
  {Gubser}}, \bibinfo {author} {\bibfnamefont {D.~R.}\ \bibnamefont {Gulotta}},
  \bibinfo {author} {\bibfnamefont {S.~S.}\ \bibnamefont {Pufu}}, \ and\
  \bibinfo {author} {\bibfnamefont {F.~D.}\ \bibnamefont {Rocha}},\ }\href
  {\doibase 10.1088/1126-6708/2008/10/052} {\bibfield  {journal} {\bibinfo
  {journal} {JHEP}\ }\textbf {\bibinfo {volume} {0810}},\ \bibinfo {pages}
  {052} (\bibinfo {year} {2008}{\natexlab{a}})},\ \Eprint
  {http://arxiv.org/abs/0803.1470} {arXiv:0803.1470 [hep-th]} \BibitemShut
  {NoStop}%
\bibitem [{\citenamefont {Chesler}\ \emph
  {et~al.}(2009{\natexlab{a}})\citenamefont {Chesler}, \citenamefont {Jensen},\
  and\ \citenamefont {Karch}}]{Chesler:2008wd}%
  \BibitemOpen
  \bibfield  {author} {\bibinfo {author} {\bibfnamefont {P.~M.}\ \bibnamefont
  {Chesler}}, \bibinfo {author} {\bibfnamefont {K.}~\bibnamefont {Jensen}}, \
  and\ \bibinfo {author} {\bibfnamefont {A.}~\bibnamefont {Karch}},\ }\href
  {\doibase 10.1103/PhysRevD.79.025021} {\bibfield  {journal} {\bibinfo
  {journal} {Phys. Rev.}\ }\textbf {\bibinfo {volume} {D79}},\ \bibinfo {pages}
  {025021} (\bibinfo {year} {2009}{\natexlab{a}})},\ \Eprint
  {http://arxiv.org/abs/0804.3110} {arXiv:0804.3110 [hep-th]} \BibitemShut
  {NoStop}%
\bibitem [{\citenamefont {Hatta}\ \emph {et~al.}(2008)\citenamefont {Hatta},
  \citenamefont {Iancu},\ and\ \citenamefont {Mueller}}]{Hatta:2008tx}%
  \BibitemOpen
  \bibfield  {author} {\bibinfo {author} {\bibfnamefont {Y.}~\bibnamefont
  {Hatta}}, \bibinfo {author} {\bibfnamefont {E.}~\bibnamefont {Iancu}}, \ and\
  \bibinfo {author} {\bibfnamefont {A.~H.}\ \bibnamefont {Mueller}},\ }\href
  {\doibase 10.1088/1126-6708/2008/05/037} {\bibfield  {journal} {\bibinfo
  {journal} {JHEP}\ }\textbf {\bibinfo {volume} {05}},\ \bibinfo {pages} {037}
  (\bibinfo {year} {2008})},\ \Eprint {http://arxiv.org/abs/0803.2481}
  {arXiv:0803.2481 [hep-th]} \BibitemShut {NoStop}%
\bibitem [{\citenamefont {Ficnar}(2012)}]{f12}%
  \BibitemOpen
  \bibfield  {author} {\bibinfo {author} {\bibfnamefont {A.}~\bibnamefont
  {Ficnar}},\ }\href {\doibase 10.1103/PhysRevD.86.046010} {\bibfield
  {journal} {\bibinfo  {journal} {Phys.Rev.}\ }\textbf {\bibinfo {volume}
  {D86}},\ \bibinfo {pages} {046010} (\bibinfo {year} {2012})},\ \Eprint
  {http://arxiv.org/abs/1201.1780} {arXiv:1201.1780 [hep-th]} \BibitemShut
  {NoStop}%
\bibitem [{\citenamefont {Chesler}\ \emph
  {et~al.}(2009{\natexlab{b}})\citenamefont {Chesler}, \citenamefont {Jensen},
  \citenamefont {Karch},\ and\ \citenamefont {Yaffe}}]{cjky08}%
  \BibitemOpen
  \bibfield  {author} {\bibinfo {author} {\bibfnamefont {P.~M.}\ \bibnamefont
  {Chesler}}, \bibinfo {author} {\bibfnamefont {K.}~\bibnamefont {Jensen}},
  \bibinfo {author} {\bibfnamefont {A.}~\bibnamefont {Karch}}, \ and\ \bibinfo
  {author} {\bibfnamefont {L.~G.}\ \bibnamefont {Yaffe}},\ }\href {\doibase
  10.1103/PhysRevD.79.125015} {\bibfield  {journal} {\bibinfo  {journal}
  {Phys.Rev.}\ }\textbf {\bibinfo {volume} {D79}},\ \bibinfo {pages} {125015}
  (\bibinfo {year} {2009}{\natexlab{b}})},\ \Eprint
  {http://arxiv.org/abs/0810.1985} {arXiv:0810.1985 [hep-th]} \BibitemShut
  {NoStop}%
\bibitem [{\citenamefont {Ficnar}\ \emph {et~al.}(2013)\citenamefont {Ficnar},
  \citenamefont {Noronha},\ and\ \citenamefont {Gyulassy}}]{fng12}%
  \BibitemOpen
  \bibfield  {author} {\bibinfo {author} {\bibfnamefont {A.}~\bibnamefont
  {Ficnar}}, \bibinfo {author} {\bibfnamefont {J.}~\bibnamefont {Noronha}}, \
  and\ \bibinfo {author} {\bibfnamefont {M.}~\bibnamefont {Gyulassy}},\
  }\href@noop {} {\bibfield  {journal} {\bibinfo  {journal} {Nucl. Phys.}\
  }\textbf {\bibinfo {volume} {A910-911}},\ \bibinfo {pages} {252} (\bibinfo
  {year} {2013})},\ \Eprint {http://arxiv.org/abs/1208.0305} {arXiv:1208.0305
  [hep-ph]} \BibitemShut {NoStop}%
\bibitem [{\citenamefont {Ficnar}\ and\ \citenamefont {Gubser}(2014)}]{fg13}%
  \BibitemOpen
  \bibfield  {author} {\bibinfo {author} {\bibfnamefont {A.}~\bibnamefont
  {Ficnar}}\ and\ \bibinfo {author} {\bibfnamefont {S.~S.}\ \bibnamefont
  {Gubser}},\ }\href {\doibase 10.1103/PhysRevD.89.026002} {\bibfield
  {journal} {\bibinfo  {journal} {Phys.Rev.}\ }\textbf {\bibinfo {volume}
  {D89}},\ \bibinfo {pages} {026002} (\bibinfo {year} {2014})},\ \Eprint
  {http://arxiv.org/abs/1306.6648} {arXiv:1306.6648 [hep-th]} \BibitemShut
  {NoStop}%
\bibitem [{\citenamefont {Chatrchyan}\ \emph
  {et~al.}(2012{\natexlab{a}})\citenamefont {Chatrchyan} \emph
  {et~al.}}]{cms12}%
  \BibitemOpen
  \bibfield  {author} {\bibinfo {author} {\bibfnamefont {S.}~\bibnamefont
  {Chatrchyan}} \emph {et~al.} (\bibinfo {collaboration} {CMS Collaboration}),\
  }\href {\doibase 10.1140/epjc/s10052-012-1945-x} {\bibfield  {journal}
  {\bibinfo  {journal} {Eur.Phys.J.}\ }\textbf {\bibinfo {volume} {C72}},\
  \bibinfo {pages} {1945} (\bibinfo {year} {2012}{\natexlab{a}})},\ \Eprint
  {http://arxiv.org/abs/1202.2554} {arXiv:1202.2554 [nucl-ex]} \BibitemShut
  {NoStop}%
\bibitem [{\citenamefont {Guijosa}\ and\ \citenamefont {Pedraza}(2011)}]{gp11}%
  \BibitemOpen
  \bibfield  {author} {\bibinfo {author} {\bibfnamefont {A.}~\bibnamefont
  {Guijosa}}\ and\ \bibinfo {author} {\bibfnamefont {J.~F.}\ \bibnamefont
  {Pedraza}},\ }\href {\doibase 10.1007/JHEP05(2011)108} {\bibfield  {journal}
  {\bibinfo  {journal} {JHEP}\ }\textbf {\bibinfo {volume} {1105}},\ \bibinfo
  {pages} {108} (\bibinfo {year} {2011})},\ \Eprint
  {http://arxiv.org/abs/1102.4893} {arXiv:1102.4893 [hep-th]} \BibitemShut
  {NoStop}%
\bibitem [{\citenamefont {Horowitz}\ and\ \citenamefont
  {Gyulassy}(2011)}]{hg11}%
  \BibitemOpen
  \bibfield  {author} {\bibinfo {author} {\bibfnamefont {W.}~\bibnamefont
  {Horowitz}}\ and\ \bibinfo {author} {\bibfnamefont {M.}~\bibnamefont
  {Gyulassy}},\ }\href {\doibase 10.1016/j.nuclphysa.2011.09.018} {\bibfield
  {journal} {\bibinfo  {journal} {Nucl.Phys.}\ }\textbf {\bibinfo {volume}
  {A872}},\ \bibinfo {pages} {265} (\bibinfo {year} {2011})},\ \Eprint
  {http://arxiv.org/abs/1104.4958} {arXiv:1104.4958 [hep-ph]} \BibitemShut
  {NoStop}%
\bibitem [{\citenamefont {Noronha}\ \emph {et~al.}(2010)\citenamefont
  {Noronha}, \citenamefont {Gyulassy},\ and\ \citenamefont {Torrieri}}]{ngt10}%
  \BibitemOpen
  \bibfield  {author} {\bibinfo {author} {\bibfnamefont {J.}~\bibnamefont
  {Noronha}}, \bibinfo {author} {\bibfnamefont {M.}~\bibnamefont {Gyulassy}}, \
  and\ \bibinfo {author} {\bibfnamefont {G.}~\bibnamefont {Torrieri}},\ }\href
  {\doibase 10.1103/PhysRevC.82.054903} {\bibfield  {journal} {\bibinfo
  {journal} {Phys.Rev.}\ }\textbf {\bibinfo {volume} {C82}},\ \bibinfo {pages}
  {054903} (\bibinfo {year} {2010})},\ \Eprint {http://arxiv.org/abs/1009.2286}
  {arXiv:1009.2286 [nucl-th]} \BibitemShut {NoStop}%
\bibitem [{\citenamefont {Gubser}(2007)}]{g07}%
  \BibitemOpen
  \bibfield  {author} {\bibinfo {author} {\bibfnamefont {S.~S.}\ \bibnamefont
  {Gubser}},\ }\href {\doibase 10.1103/PhysRevD.76.126003} {\bibfield
  {journal} {\bibinfo  {journal} {Phys.Rev.}\ }\textbf {\bibinfo {volume}
  {D76}},\ \bibinfo {pages} {126003} (\bibinfo {year} {2007})},\ \Eprint
  {http://arxiv.org/abs/hep-th/0611272} {arXiv:hep-th/0611272 [hep-th]}
  \BibitemShut {NoStop}%
\bibitem [{\citenamefont {Betz}\ and\ \citenamefont {Gyulassy}(2013)}]{bg13}%
  \BibitemOpen
  \bibfield  {author} {\bibinfo {author} {\bibfnamefont {B.}~\bibnamefont
  {Betz}}\ and\ \bibinfo {author} {\bibfnamefont {M.}~\bibnamefont
  {Gyulassy}},\ }\href@noop {} {\  (\bibinfo {year} {2013})},\ \Eprint
  {http://arxiv.org/abs/1305.6458} {arXiv:1305.6458 [nucl-th]} \BibitemShut
  {NoStop}%
\bibitem [{\citenamefont {Kniehl}\ \emph {et~al.}(2000)\citenamefont {Kniehl},
  \citenamefont {Kramer},\ and\ \citenamefont {Potter}}]{kkp00}%
  \BibitemOpen
  \bibfield  {author} {\bibinfo {author} {\bibfnamefont {B.~A.}\ \bibnamefont
  {Kniehl}}, \bibinfo {author} {\bibfnamefont {G.}~\bibnamefont {Kramer}}, \
  and\ \bibinfo {author} {\bibfnamefont {B.}~\bibnamefont {Potter}},\ }\href
  {\doibase 10.1016/S0550-3213(00)00303-5} {\bibfield  {journal} {\bibinfo
  {journal} {Nucl.Phys.}\ }\textbf {\bibinfo {volume} {B582}},\ \bibinfo
  {pages} {514} (\bibinfo {year} {2000})},\ \Eprint
  {http://arxiv.org/abs/hep-ph/0010289} {arXiv:hep-ph/0010289 [hep-ph]}
  \BibitemShut {NoStop}%
\bibitem [{\citenamefont {Xu}\ \emph {et~al.}(2014)\citenamefont {Xu},
  \citenamefont {Buzzatti},\ and\ \citenamefont {Gyulassy}}]{JiechenCUJET2}%
  \BibitemOpen
  \bibfield  {author} {\bibinfo {author} {\bibfnamefont {J.}~\bibnamefont
  {Xu}}, \bibinfo {author} {\bibfnamefont {A.}~\bibnamefont {Buzzatti}}, \ and\
  \bibinfo {author} {\bibfnamefont {M.}~\bibnamefont {Gyulassy}},\ }\href@noop
  {} {\  (\bibinfo {year} {2014})},\ \Eprint {http://arxiv.org/abs/1402.2956}
  {arXiv:1402.2956 [hep-ph]} \BibitemShut {NoStop}%
\bibitem [{\citenamefont {Wang}()}]{WangPrivate}%
  \BibitemOpen
  \bibfield  {author} {\bibinfo {author} {\bibfnamefont {X.-N.}\ \bibnamefont
  {Wang}},\ }\href@noop {} {\enquote {\bibinfo {title} {{Private
  communication}},}\ }\BibitemShut {NoStop}%
\bibitem [{\citenamefont {Deng}\ \emph {et~al.}(2011)\citenamefont {Deng},
  \citenamefont {Wang},\ and\ \citenamefont {Xu}}]{dwx11}%
  \BibitemOpen
  \bibfield  {author} {\bibinfo {author} {\bibfnamefont {W.-T.}\ \bibnamefont
  {Deng}}, \bibinfo {author} {\bibfnamefont {X.-N.}\ \bibnamefont {Wang}}, \
  and\ \bibinfo {author} {\bibfnamefont {R.}~\bibnamefont {Xu}},\ }\href
  {\doibase 10.1103/PhysRevC.83.014915} {\bibfield  {journal} {\bibinfo
  {journal} {Phys.Rev.}\ }\textbf {\bibinfo {volume} {C83}},\ \bibinfo {pages}
  {014915} (\bibinfo {year} {2011})},\ \Eprint {http://arxiv.org/abs/1008.1841}
  {arXiv:1008.1841 [hep-ph]} \BibitemShut {NoStop}%
\bibitem [{\citenamefont {Gubser}\ \emph
  {et~al.}(2008{\natexlab{b}})\citenamefont {Gubser}, \citenamefont {Nellore},
  \citenamefont {Pufu},\ and\ \citenamefont {Rocha}}]{gnpr08}%
  \BibitemOpen
  \bibfield  {author} {\bibinfo {author} {\bibfnamefont {S.~S.}\ \bibnamefont
  {Gubser}}, \bibinfo {author} {\bibfnamefont {A.}~\bibnamefont {Nellore}},
  \bibinfo {author} {\bibfnamefont {S.~S.}\ \bibnamefont {Pufu}}, \ and\
  \bibinfo {author} {\bibfnamefont {F.~D.}\ \bibnamefont {Rocha}},\ }\href
  {\doibase 10.1103/PhysRevLett.101.131601} {\bibfield  {journal} {\bibinfo
  {journal} {Phys.Rev.Lett.}\ }\textbf {\bibinfo {volume} {101}},\ \bibinfo
  {pages} {131601} (\bibinfo {year} {2008}{\natexlab{b}})},\ \Eprint
  {http://arxiv.org/abs/0804.1950} {arXiv:0804.1950 [hep-th]} \BibitemShut
  {NoStop}%
\bibitem [{\citenamefont {Ficnar}\ \emph {et~al.}(2011)\citenamefont {Ficnar},
  \citenamefont {Noronha},\ and\ \citenamefont {Gyulassy}}]{fng11}%
  \BibitemOpen
  \bibfield  {author} {\bibinfo {author} {\bibfnamefont {A.}~\bibnamefont
  {Ficnar}}, \bibinfo {author} {\bibfnamefont {J.}~\bibnamefont {Noronha}}, \
  and\ \bibinfo {author} {\bibfnamefont {M.}~\bibnamefont {Gyulassy}},\ }\href
  {\doibase 10.1088/0954-3899/38/12/124176} {\bibfield  {journal} {\bibinfo
  {journal} {J.Phys.}\ }\textbf {\bibinfo {volume} {G38}},\ \bibinfo {pages}
  {124176} (\bibinfo {year} {2011})},\ \Eprint {http://arxiv.org/abs/1106.6303}
  {arXiv:1106.6303 [hep-ph]} \BibitemShut {NoStop}%
\bibitem [{\citenamefont {Adare}\ \emph {et~al.}(2012)\citenamefont {Adare}
  \emph {et~al.}}]{phenix12}%
  \BibitemOpen
  \bibfield  {author} {\bibinfo {author} {\bibfnamefont {A.}~\bibnamefont
  {Adare}} \emph {et~al.} (\bibinfo {collaboration} {PHENIX Collaboration}),\
  }\href@noop {} {\  (\bibinfo {year} {2012})},\ \Eprint
  {http://arxiv.org/abs/1208.2254} {arXiv:1208.2254 [nucl-ex]} \BibitemShut
  {NoStop}%
\bibitem [{\citenamefont {Fadafan}(2008)}]{f08}%
  \BibitemOpen
  \bibfield  {author} {\bibinfo {author} {\bibfnamefont {K.~B.}\ \bibnamefont
  {Fadafan}},\ }\href {\doibase 10.1088/1126-6708/2008/12/051} {\bibfield
  {journal} {\bibinfo  {journal} {JHEP}\ }\textbf {\bibinfo {volume} {0812}},\
  \bibinfo {pages} {051} (\bibinfo {year} {2008})},\ \Eprint
  {http://arxiv.org/abs/0803.2777} {arXiv:0803.2777 [hep-th]} \BibitemShut
  {NoStop}%
\bibitem [{\citenamefont {Brigante}\ \emph {et~al.}(2008)\citenamefont
  {Brigante}, \citenamefont {Liu}, \citenamefont {Myers}, \citenamefont
  {Shenker},\ and\ \citenamefont {Yaida}}]{blmsy08}%
  \BibitemOpen
  \bibfield  {author} {\bibinfo {author} {\bibfnamefont {M.}~\bibnamefont
  {Brigante}}, \bibinfo {author} {\bibfnamefont {H.}~\bibnamefont {Liu}},
  \bibinfo {author} {\bibfnamefont {R.~C.}\ \bibnamefont {Myers}}, \bibinfo
  {author} {\bibfnamefont {S.}~\bibnamefont {Shenker}}, \ and\ \bibinfo
  {author} {\bibfnamefont {S.}~\bibnamefont {Yaida}},\ }\href {\doibase
  10.1103/PhysRevD.77.126006} {\bibfield  {journal} {\bibinfo  {journal}
  {Phys.Rev.}\ }\textbf {\bibinfo {volume} {D77}},\ \bibinfo {pages} {126006}
  (\bibinfo {year} {2008})},\ \Eprint {http://arxiv.org/abs/0712.0805}
  {arXiv:0712.0805 [hep-th]} \BibitemShut {NoStop}%
\bibitem [{\citenamefont {Hofman}\ and\ \citenamefont
  {Maldacena}(2008)}]{hm08}%
  \BibitemOpen
  \bibfield  {author} {\bibinfo {author} {\bibfnamefont {D.~M.}\ \bibnamefont
  {Hofman}}\ and\ \bibinfo {author} {\bibfnamefont {J.}~\bibnamefont
  {Maldacena}},\ }\href {\doibase 10.1088/1126-6708/2008/05/012} {\bibfield
  {journal} {\bibinfo  {journal} {JHEP}\ }\textbf {\bibinfo {volume} {0805}},\
  \bibinfo {pages} {012} (\bibinfo {year} {2008})},\ \Eprint
  {http://arxiv.org/abs/0803.1467} {arXiv:0803.1467 [hep-th]} \BibitemShut
  {NoStop}%
\bibitem [{\citenamefont {Cai}(2002)}]{c02}%
  \BibitemOpen
  \bibfield  {author} {\bibinfo {author} {\bibfnamefont {R.-G.}\ \bibnamefont
  {Cai}},\ }\href {\doibase 10.1103/PhysRevD.65.084014} {\bibfield  {journal}
  {\bibinfo  {journal} {Phys.Rev.}\ }\textbf {\bibinfo {volume} {D65}},\
  \bibinfo {pages} {084014} (\bibinfo {year} {2002})},\ \Eprint
  {http://arxiv.org/abs/hep-th/0109133} {arXiv:hep-th/0109133 [hep-th]}
  \BibitemShut {NoStop}%
\bibitem [{\citenamefont {Song}\ \emph {et~al.}(2013)\citenamefont {Song},
  \citenamefont {Bass},\ and\ \citenamefont {Heinz}}]{sbh13}%
  \BibitemOpen
  \bibfield  {author} {\bibinfo {author} {\bibfnamefont {H.}~\bibnamefont
  {Song}}, \bibinfo {author} {\bibfnamefont {S.}~\bibnamefont {Bass}}, \ and\
  \bibinfo {author} {\bibfnamefont {U.~W.}\ \bibnamefont {Heinz}},\ }\href@noop
  {} {\  (\bibinfo {year} {2013})},\ \Eprint {http://arxiv.org/abs/1311.0157}
  {arXiv:1311.0157 [nucl-th]} \BibitemShut {NoStop}%
\bibitem [{\citenamefont {Chatrchyan}\ \emph
  {et~al.}(2012{\natexlab{b}})\citenamefont {Chatrchyan} \emph
  {et~al.}}]{cms12a}%
  \BibitemOpen
  \bibfield  {author} {\bibinfo {author} {\bibfnamefont {S.}~\bibnamefont
  {Chatrchyan}} \emph {et~al.} (\bibinfo {collaboration} {CMS Collaboration}),\
  }\href {\doibase 10.1103/PhysRevLett.109.022301} {\bibfield  {journal}
  {\bibinfo  {journal} {Phys.Rev.Lett.}\ }\textbf {\bibinfo {volume} {109}},\
  \bibinfo {pages} {022301} (\bibinfo {year} {2012}{\natexlab{b}})},\ \Eprint
  {http://arxiv.org/abs/1204.1850} {arXiv:1204.1850 [nucl-ex]} \BibitemShut
  {NoStop}%
\bibitem [{\citenamefont {Song}\ \emph {et~al.}(2011)\citenamefont {Song},
  \citenamefont {Bass}, \citenamefont {Heinz}, \citenamefont {Hirano},\ and\
  \citenamefont {Shen}}]{sbhhs11}%
  \BibitemOpen
  \bibfield  {author} {\bibinfo {author} {\bibfnamefont {H.}~\bibnamefont
  {Song}}, \bibinfo {author} {\bibfnamefont {S.~A.}\ \bibnamefont {Bass}},
  \bibinfo {author} {\bibfnamefont {U.}~\bibnamefont {Heinz}}, \bibinfo
  {author} {\bibfnamefont {T.}~\bibnamefont {Hirano}}, \ and\ \bibinfo {author}
  {\bibfnamefont {C.}~\bibnamefont {Shen}},\ }\href {\doibase
  10.1103/PhysRevC.83.054910, 10.1103/PhysRevC.86.059903} {\bibfield  {journal}
  {\bibinfo  {journal} {Phys.Rev.}\ }\textbf {\bibinfo {volume} {C83}},\
  \bibinfo {pages} {054910} (\bibinfo {year} {2011})},\ \Eprint
  {http://arxiv.org/abs/1101.4638} {arXiv:1101.4638 [nucl-th]} \BibitemShut
  {NoStop}%
\bibitem [{\citenamefont {Gubser}(2006)}]{g06}%
  \BibitemOpen
  \bibfield  {author} {\bibinfo {author} {\bibfnamefont {S.~S.}\ \bibnamefont
  {Gubser}},\ }\href {\doibase 10.1103/PhysRevD.74.126005} {\bibfield
  {journal} {\bibinfo  {journal} {Phys.Rev.}\ }\textbf {\bibinfo {volume}
  {D74}},\ \bibinfo {pages} {126005} (\bibinfo {year} {2006})},\ \Eprint
  {http://arxiv.org/abs/hep-th/0605182} {arXiv:hep-th/0605182 [hep-th]}
  \BibitemShut {NoStop}%
\bibitem [{\citenamefont {Herzog}\ \emph {et~al.}(2006)\citenamefont {Herzog},
  \citenamefont {Karch}, \citenamefont {Kovtun}, \citenamefont {Kozcaz},\ and\
  \citenamefont {Yaffe}}]{hkkky06}%
  \BibitemOpen
  \bibfield  {author} {\bibinfo {author} {\bibfnamefont {C.}~\bibnamefont
  {Herzog}}, \bibinfo {author} {\bibfnamefont {A.}~\bibnamefont {Karch}},
  \bibinfo {author} {\bibfnamefont {P.}~\bibnamefont {Kovtun}}, \bibinfo
  {author} {\bibfnamefont {C.}~\bibnamefont {Kozcaz}}, \ and\ \bibinfo {author}
  {\bibfnamefont {L.}~\bibnamefont {Yaffe}},\ }\href {\doibase
  10.1088/1126-6708/2006/07/013} {\bibfield  {journal} {\bibinfo  {journal}
  {JHEP}\ }\textbf {\bibinfo {volume} {0607}},\ \bibinfo {pages} {013}
  (\bibinfo {year} {2006})},\ \Eprint {http://arxiv.org/abs/hep-th/0605158}
  {arXiv:hep-th/0605158 [hep-th]} \BibitemShut {NoStop}%
\end{thebibliography}%


\end{document}